\documentclass[aps,prb,twocolumn,amsmath,amssymb,superscriptaddress,floatfix]{revtex4-1}

\usepackage{graphicx}
\usepackage{multirow}
\usepackage{color}
\usepackage{bm}

\renewcommand{\Vec}[1]{\mbox{\boldmath$#1$}}
\def\infinity{\infty}
\def\t#1{\textrm{#1}}
\def\ket#1{|#1\rangle }
\def\bra#1{\langle #1 |}
\def\n{\nonumber \\ }
\def\tensor{\otimes}

\newcommand{\overbar}[1]{\mkern 1.5mu\overline{\mkern-1.5mu#1\mkern-1.5mu}\mkern 1.5mu}

\begin{document}

\title{
$\mathbb{Z}_3$ symmetry-protected topological phases
in the SU(3) AKLT model
}

\author{Takahiro Morimoto}
\affiliation{Condensed Matter Theory Laboratory, RIKEN, Wako, Saitama, 351-0198, Japan}
\author{Hiroshi Ueda}
\affiliation{Condensed Matter Theory Laboratory, RIKEN, Wako, Saitama, 351-0198, Japan}
\author{Tsutomu Momoi}
\affiliation{Condensed Matter Theory Laboratory, RIKEN, Wako, Saitama, 351-0198, Japan}
\affiliation{RIKEN Center for Emergent Matter Science (CEMS), Wako, Saitama, 351-0198, Japan}
\author{Akira Furusaki}
\affiliation{Condensed Matter Theory Laboratory, RIKEN, Wako, Saitama, 351-0198, Japan}
\affiliation{RIKEN Center for Emergent Matter Science (CEMS), Wako, Saitama, 351-0198, Japan}

\date{\today}

\begin{abstract}
We study $\mathbb{Z}_3$ symmetry-protected topological (SPT) phases
in one-dimensional spin systems with $Z_3 \times Z_3$ symmetry.
We construct ground-state wave functions of the matrix product form
for nontrivial $\mathbb{Z}_3$ phases and their parent Hamiltonian
from a cocycle of the group cohomology $H^2(Z_3\times Z_3,U(1))$.
The Hamiltonian is an SU(3) version of the
Affleck-Kennedy-Lieb-Tasaki (AKLT) model,
consisting of bilinear and biquadratic terms of
su(3) generators in the adjoint representation.
A generalization to the SU($N$) case, the SU($N$) AKLT Hamiltonian, is
also presented which realizes nontrivial $\mathbb{Z}_N$ SPT phases.
We use the infinite-size variant of the density matrix renormalization
group (iDMRG) method to
determine the ground-state phase diagram of
the SU(3) bilinear-biquadratic model as a function of the parameter $\theta$
controlling the ratio of the bilinear and biquadratic coupling constants.
The nontrivial $\mathbb{Z}_3$ SPT phase is found
for a range of the parameter $\theta$ including the point of vanishing
biquadratic term ($\theta=0$) as well as the SU(3) AKLT point
[$\theta=\arctan(2/9)$].
A continuous phase transition to the SU(3) dimer phase takes place
at $\theta \approx  -0.027\pi$, with a central charge $c\approx3.2$.
For SU(3) symmetric cases we define string order parameters
for the $\mathbb{Z}_3$ SPT phases in a similar way
to the conventional Haldane phase.
We propose simple spin models that effectively realize
the SU(3) and SU(4) AKLT models.
\end{abstract}

\maketitle

\section{Introduction}

The Haldane phase\cite{HaldanePL1983,HaldanePRL1983}
of antiferromagnetic $S=1$ spin chains
is a representative topological phase of one-dimensional (1D) gapped
quantum systems.
In the Haldane phase, excitations are gapped in the bulk,
while zero-energy states of effective $S=1/2$ spins
are present at the boundaries.
The essence of the Haldane phase is captured by the toy model
proposed by Affleck, Kennedy, Lieb, and Tasaki (AKLT),\cite{AKLT-prl87,AKLT-cmp88}
which is constructed from projection operators acting on two neighboring sites.
Its ground state (the AKLT state) has the following structure.
Each $S=1$ spin is decomposed into
two virtual $S=1/2$ spins.
On each site two $S=1/2$ spins are symmetrized to
form an $S=1$ spin,
while two $S=1/2$ spins from neighboring sites form
a singlet on each bond.
At each end of the spin chain, an effective $S=1/2$ spin
is left without forming a singlet and realizes
two-fold degenerate zero modes.
The AKLT state shows no apparent symmetry breaking such as
magnetic order and lattice symmetry breaking.
However, it has a hidden order called the string order,\cite{denNijsRommelse}
which corresponds to a ferromagnetic order
in the system after
a non-local unitary transformation.\cite{KennedyTasaki,KennedyTasaki2}
The string order signals a hidden $Z_2\times Z_2$ symmetry breaking
in the Haldane phase.


Recent advances in the understanding of 1D topological phases
are brought by the notion of
symmetry protected topological (SPT) phases.\cite{pollmann10,pollmann11,fidkowski-kitaev11,chen-gu-wen11,chen-science12}
The Haldane phase is an SPT phase that is protected by
any one of the following symmetries:\cite{pollmann11}
(a) time-reversal symmetry, (b) link inversion symmetry, and
(c) the dihedral group of $\pi$ rotations about the
$S_x$, $S_y$, and $S_z$ axes.
Here let us assume the $Z_2 \times Z_2$ symmetry of the dihedral group.
The AKLT Hamiltonian is invariant under the $\pi$ rotation around
the $S_x$ and $S_z$ axes, and these $\pi$ rotations commute
with each other for the original $S=1$ spins.
However, they do not commute (in fact anticommute)
with each other for the virtual $S=1/2$ spins.
This is an example of projective representations of symmetry groups,
i.e., symmetry operations represented projectively
on the effective (fractionalized) degrees of freedom
which appear at the boundaries.
This can be nicely formulated in the framework of
matrix product states (MPSs) for 1D gapped systems.
The AKLT wave function is written in the MPS form
with $2\times2$ matrices acting on the two states
$\ket{\mbox{$\uparrow$}}$, $\ket{\mbox{$\downarrow$}}$
of a virtual $S=1/2$ spin.
Symmetry operations ($\pi$ rotations) acting on the three states
of each $S=1$ spin induce linear transformations of the $2\times2$
matrices, which are then expressed as unitary transformations
in the two-dimensional space spanned by
$\ket{\mbox{$\uparrow$}}$ and $\ket{\mbox{$\downarrow$}}$.
The unitary matrices of this basis transformation
give a projective representation of the symmetry group
with a phase factor which is
an element of the group cohomology $H^2(Z_2 \times Z_2,U(1))$.
The Haldane phase is an example of SPT phases
and corresponds to the nontrivial element of
$H^2(Z_2 \times Z_2,U(1))=\mathbb{Z}_2$.
In general 1D SPT phases protected by symmetry group $G$ are classified
in terms of the second cohomology group $H^2(G,U(1))$
of the group $G$.\cite{chen-gu-wen11,fidkowski-kitaev11,schuch11,chen-science12}


In this paper we generalize the AKLT state of the Haldane phase
to 1D SPT phases protected by $Z_N\times Z_N$ symmetry.
We focus on the case of $N=3$ and briefly discuss the general case $N>3$.
Our starting point is the observation that
$Z_3\times Z_3$ symmetry can be projectively represented
by $3\times3$ matrices, with a U(1) phase factor which is
a nontrivial element of
$H^2(Z_3 \times Z_3,U(1))=\mathbb{Z}_3$.
This observation allows us to write down MPS wave functions
with $3\times3$ matrices as described below,
as a natural generalization of the AKLT state.
The MPS wave functions are ground states of an SU(3) generalization
of the AKLT model and describe topological states in
$\mathbb{Z}_3$ SPT phases.

We construct the SU(3) AKLT states on a 1D lattice
where the local Hilbert space on each site is spanned by
eight states of the adjoint representation $\bm 8$ of su(3),
which we call meson states.
The eight meson states are represented by traceless bilinear forms
of two sets of three virtual degrees of freedom,
i.e., three quarks ($u,d,s$) in the fundamental representation $\bm 3$
and three antiquarks ($\bar u,\bar d, \bar s$) in the conjugate
representation $\bar{\bm3}$.
The SU(3) AKLT states are valence bond solids in which
a quark and an antiquark on neighboring sites form a singlet
state on the bond connecting the two sites,
whereas a quark and an antiquark on the same site
form a meson state.
When the 1D chain has ends,
three-fold degenerate boundary zeromodes appear at each end,
which are either unpaired quark or antiquark states.
The possibility of having two types (quark or antiquark) of zeromodes
indicates that there are two distinct types of SU(3) AKLT states,
each of which represents a distinct $\mathbb{Z}_3$ SPT phase.
Both SU(3) AKLT states are ground states of the SU(3) AKLT Hamiltonian
which consists of bilinear and biquadratic terms of su(3)
generators in the $\bm8$ representation with a particular ratio of the
two terms.
The SU(3) Hamiltonian and its ground-state wave functions were in fact
presented earlier in Refs.~\onlinecite{greiter-su3-07,greiter2-07,katsura08}.
In this paper we characterize the SU(3) AKLT states as $\mathbb{Z}_3$
SPT states in the classification in terms of group cohomology 
$H^2(Z_3 \times Z_3,U(1))=\mathbb{Z}_3$
and report results of detailed study on their correlation functions and
a quantum phase transition to a dimerized phase.
We note that Refs.~\onlinecite{KD-quella-2-12,KD-quella-1-13} studied 
PSU(3) symmetric spin chains which realize $\mathbb{Z}_3$
SPT phases corresponding to nontrivial elements of
$H^3(\mbox{PSU(3),U(1)})=\mathbb{Z}_3$.
The SU(3) AKLT Hamiltonian can also be considered as a PSU(3) symmetric
model realizing $\mathbb{Z}_3$ SPT phases protected by PSU(3) symmetry
in that the adjoint representation of SU(3) is also a representation of PSU(3).

We can further generalize the SU(3) AKLT Hamiltonian
to the SU($N$) AKLT Hamiltonian ($N>3$)
consisting of bilinear and biquadratic terms of
the su($N$) generators in the adjoint representation $\bm{N^2-1}$.
Its two-fold degenerate ground state (under periodic boundary conditions)
is given by SU($N$) AKLT states which are MPSs with $N\times N$ matrices.
The SU($N$) AKLT states are valence bond solids in which states in
the $\bm{N^2-1}$ representation are decomposed into
products of states from $\bm N$ and $\bm{\overbar{N}}$ representations, which
form $\bm{N^2-1}$ and singlet states
on each site and bond, respectively.
The SU($N$) AKLT model has an energy gap as its two-point correlation
functions of SU($N$) operators are short-ranged with a correlation
length being equal to $\xi_N=1/\ln(N^2-1)$.
Realizations of SPT phases with SU($N$) symmetry in other representations
are proposed in the context of cold atoms.\cite{nonne13}

As in the SU(2) AKLT state, the SU($N$) AKLT states have
a hidden long-range order.
To see this for the SU(3) AKLT model,
we define string order parameters that characterize the $\mathbb{Z}_3$ SPT phase
by making use of the system's full SU(3) symmetry.
Similar to the conventional string order
parameter for the SU(2) AKLT state
which indicates the antiferromagnetic order
upon neglecting $S_z=0$ states,
the string order parameters for the SU(3) AKLT states
have string operators from SU(3) operators (analogous to the $S_z$ operator)
which count the number of constituent quarks or antiquarks.
We show the long-range order of string correlations by
explicitly calculating string order parameters in the SU(3) AKLT states.
Incidentally, the string orders that we define are different from those studied
in Refs.~\onlinecite{KD-quella-1-13,KD-quella-2-12,KD-quella-3-13}
where only $Z_3\times Z_3$ symmetry is assumed.

As the ratio of the two coupling constants in the SU(3) AKLT Hamiltonian
is varied,
a quantum phase transition occurs
from a $\mathbb{Z}_3$ SPT phase to a topologically trivial dimer phase
which breaks translation symmetry.
We study this topological phase transition using 
the infinite-size variant of the density matrix renormalization group (iDMRG)
method.\cite{white-92,white-93,McCulloch-08}
We obtain the phase diagram of the SU(3) bilinear-biquadratic model
and determine the location of the critical point numerically.
We find that the $\mathbb{Z}_3$ SPT phase occupies a finite region in
the parameter space and survives even when the biquadratic term is absent.
From scaling of entanglement entropy we obtain numerical evidence that
the critical point is described by the level-2 SU(3) Wess-Zumino-Witten
theory.


Finally, we demonstrate that the SU(3) AKLT Hamiltonian is realized by
an $S=1$ spin chain with staggered quadrupole couplings 
in the strong-coupling limit.
Using the fact that spin dipole and quadrupole operators of $S=1$ spins
together form eight generators of su(3)
in the fundamental representation $\bm 3$,
we construct Hamiltonians with staggered nearest-neighbor couplings of
quadrupole operators whose ground states are
smoothly connected to the SU(3) AKLT states
in the limit where positive quadrupole couplings are very strong.
In a similar manner, we propose that the SU(4) AKLT Hamiltonian
is effectively realized in the strong-coupling limit of
an $S=1/2$ spin-orbital model
which is a variant of the Kugel-Khomskii model.\cite{kugel-khomskii82}

The paper is organized as follows.
In Sec.~\ref{sec:MPS and group cohomology}
we review the MPS representation of gapped 1D quantum systems
and the classification of 1D SPT phases in terms of group cohomology.
In Sec.~\ref{sec: SU(3) AKLT model}
we construct the SU(3) AKLT model from a nontrivial cocycle
of $H^2(Z_3 \times Z_3,U(1))$
and discuss its generalization to SU($N$).
In Sec.~\ref{sec: string order}
we define string order parameters that characterize
nontrivial $\mathbb{Z}_3$ SPT phases for
the SU(3) symmetric case.
In Sec.~\ref{sec: DMRG}
we study the SU(3) bilinear-biquadratic model
with the iDMRG method and show its ground-state phase diagram.
In Sec.~\ref{sec: SU(3) and SU(4)},
we present realizations of the SU(3) and SU(4) AKLT Hamiltonians
in an $S=1$ spin chain and an $S=1/2$ spin-orbital model.
In Sec.~\ref{sec: summary} we give a brief summary.

\section{Matrix product states and group cohomology \label{sec:MPS and group cohomology}}
In this section, we give a brief review on the classification of
the 1D SPT phases in terms of the group
cohomology\cite{chen-gu-wen11,fidkowski-kitaev11,schuch11,chen-science12}
and its application to the AKLT model for the Haldane phase.
This will serve as a basis for the generalization of the AKLT model
to the SU(3) case in the next section.

\subsection{Matrix product state}
We consider a gapped ground state of an infinite spin chain
described by a wave function $\ket{\Psi}$,
which we assume to be translation invariant.
Let us consider bipartitioning of the chain
between the site $n$ and the site $n+1$.
Then we decompose the wave function
\begin{align}
\ket{\Psi}=\sum_i w_i \ket{\psi_n^L}_i \ket{\psi_{n+1}^R}_i,
\label{SVD}
\end{align}
where $w_i$'s are singular values, and
$\ket{\psi_n^L}_i$ and $\ket{\psi_{n+1}^R}_i$ are
wave functions on the left and the right semi-infinite chains
that form orthonormal basis for the left and right Hilbert spaces.
Alternatively we can decompose the wave function between
the site $n+1$ and the site $n+2$:
\begin{align}
\ket{\Psi}=\sum_i w_i \ket{\psi_{n+1}^L}_i \ket{\psi_{n+2}^R}_i,
\end{align}
where the set of singular values are the same as in
Eq.\ (\ref{SVD}) because of the translation symmetry.
Now we write $\ket{\psi_{n+1}^L}_i$ in terms of $\ket{\psi_n^L}_i$ and
local states $\ket{m}$ at the site $n+1$ as
\begin{align}
\ket{\psi_{n+1}^L}_j= \sum_m A_{i j}^m \ket{\psi_n^L}_i \tensor \ket{m},
\end{align}
where $A^m$ is a matrix defined for each
local state $\ket{m}$
and is independent of the site $n$ 
where we cut the spin chain,
again due to the translation symmetry.

If we repeat this procedure, we can relate any two left singular vectors
$\ket{\psi_{n}^L}_j$ and $\ket{\psi_{n'}^L}_j$ with $n<n'$ as
\begin{align}
\ket{\psi_{n'}^L}_{i_{n'}}&=
\sum_{i_n,\ldots,i_{n'-1}}\sum_{m_{n+1},\ldots,m_{n'}}
A_{i_n i_{n+1}}^{m_{n+1}} \dots A_{i_{n'-1} i_{n'}}^{m_{n'}}
\nonumber\\
&\hspace*{25mm}\times
\ket{\psi_n^L}_{i_n} \tensor \ket{m_{n+1} \ldots m_{n'}}.
\end{align}
The reduced density matrix for the finite region $(n+1,\ldots,n')$
and physical quantities derived from it
can be obtained from the above equation relating
singular vectors.
If we extend this procedure to a periodic chain of length $L$,
then we obtain the MPS form of the ground-state wave function,
\begin{align}
\ket{\Psi}=\sum_{\{m_i\}}
\mathrm{tr}\!\left[
A^{m_{1}} A^{m_{2}} \dots A^{m_{L}}
\right]
\ket{m_{1} \ldots m_{L}},
\label{eq: MPS periodic}
\end{align}
where the trace is over the product of matrices $A^m$.

\subsection{Symmetry operation and MPS}
Let us suppose that the system of our interest has a symmetry
group $G$ and
its ground-state wave function $\ket{\Psi}$ is invariant
under global action of any element in $G$.
We assume that the symmetry action is local (e.g., on-site) and unitary.
Local states are transformed by action of $g\in G$ as
\begin{align}
\ket{m} \to \sum_n g_{nm}\ket{n}
\label{eq: g action on site}
\end{align}
with a unitary matrix $g_{nm}$.
The wave function $\ket{\Psi}$ is written in the form of an MPS of
Eq.~(\ref{eq: MPS periodic}),
whose transformation by $g$ is obtained
by applying Eq.~(\ref{eq: g action on site})
to the local states $\ket{m}$ at every site:
\begin{align}
\ket{\Psi}&\to
\ket{\widetilde \Psi}
\nonumber\\
&=\!\sum_{\{m_i,n_i\}}\!\mathrm{tr}\!\left[
g_{n_1 m_1} A^{m_{1}} \dots g_{n_L m_L} A^{m_{L}}
\right]\!
\ket{n_{1} \ldots n_{L}}.
\end{align}
We see that the wave function $\ket{\widetilde\Psi}$ is an MPS
made from the matrices
\begin{align}
\widetilde{A}^m=\sum_n g_{mn}A^n.
\end{align}
We demand that the ground state $\ket{\Psi}$ be invariant
up to a phase factor, i.e.,
$\ket{\widetilde\Psi}=e^{i L \theta_g}\ket{\Psi}$.
This is achieved if
\begin{align}
\sum_n g_{mn} A^n = e^{i\theta_g} U_g^{-1} A^m U_g,
\label{eq: symmetry transformation of MPS}
\end{align}
where $U_g$ is a $g$-dependent unitary matrix
which is independent of the local states $m$.
It is known that $U_g$ is unique up to a $U(1)$ phase
when the transfer matrix 
$\sum_m A^m \tensor (A^m)^*$ has only one eigenvalue of
the largest magnitude\cite{perez-garcia08,fidkowski-kitaev11}
(the state is not a macroscopic superposition of orthogonal states).

Let us consider successive actions of $g,h \in G$ on $\ket{\Psi}$,
which induce transformations
\begin{align}
\sum_{l,n} g_{ml}h_{ln} A^n
 &= e^{i\theta_g} e^{i\theta_h}  U_h^{-1} U_g^{-1} A^m U_g U_h,
\label{g*h}
\end{align}
where we have used the fact that $G$ is a unitary symmetry
(which does not include an anti-unitary operator such as time reversal),
as we assume throughout this paper.
Equation (\ref{g*h}) should coincide with the transformation
induced by an action of $gh$,
\begin{align}
\sum_n (gh)_{mn} A^n &= e^{i\theta_{gh}} U_{gh}^{-1} A^m U_{gh}.
\end{align}
We thus have
\begin{subequations}
\begin{align}
\theta_{gh}&=\theta_{g}+\theta_{h}, \\
U_g U_h&=\exp[i\phi(g,h)]U_{gh},
\label{projective representation}
\end{align}
\end{subequations}
where the second equation has a U(1) phase.
Equation (\ref{projective representation}) shows that $U_g$'s give
a projective representation of the symmetry group $G$.
The phase function $\phi(g,h)$ encodes topological data of
the ground-state wave function and
has the following two properties
[Eqs.~(\ref{eq: 2 cocycle condition}) and (\ref{eq: coboundary equivalence})]
that define group cohomology.

\textit{Cocycle:}
Let us calculate the product $U_{g_1}U_{g_2}U_{g_3}$
in two different ways (associativity):
\begin{subequations}
\begin{align}
U_{g_1} U_{g_2} U_{g_3}&=\exp[i\phi(g_2, g_3)]U_{g_1} U_{g_2 g_3} \n
&= \exp[i\phi(g_2, g_3)+i\phi(g_1,g_2 g_3)]U_{g_1 g_2 g_3}
\end{align}
and
\begin{align}
U_{g_1} U_{g_2} U_{g_3}&=\exp[i\phi(g_1, g_2)]U_{g_1 g_2} U_{g_3} \n
&= \exp[i\phi(g_1, g_2)+i\phi(g_1 g_2, g_3)]U_{g_1 g_2 g_3}.
\end{align}
\end{subequations}
The consistency between the two results requires
the phase function to satisfy
\begin{align}
\phi(g_2, g_3)-\phi(g_1 g_2, g_3)+\phi(g_1,g_2 g_3)-\phi(g_1, g_2)=0.
\label{eq: 2 cocycle condition}
\end{align}
This is the cocycle condition.
(For more mathematical details, see Appendix~\ref{app: group cohomology}.)

\textit{Coboundary:}
The ambiguity of a U(1) phase in defining a unitary matrix $U_g$
in Eq.\ (\ref{eq: symmetry transformation of MPS})
implies that we are free to take another set of unitary matrices,
\begin{align}
\widetilde U_g = \exp[i\beta(g)] U_g.
\end{align}
Accordingly, the phase function appearing in
the projective representation in Eq.\ (\ref{projective representation})
is changed from $\phi$ to $\tilde\phi$,
\begin{align}
\tilde\phi(g_1,g_2) = \phi(g_1,g_2)+[\beta(g_2)-\beta(g_1 g_2)+\beta(g_1)],
\label{eq: coboundary equivalence}
\end{align}
where the three terms in the square brackets [  ] are called 2-coboundary;
see Appendix~\ref{app: group cohomology}.
The two phase functions $\phi$ and $\tilde\phi$ are equivalent
up to a 2-coboundary and describe the same topological phase.

The set of phase functions that satisfy the cocycle condition
(\ref{eq: 2 cocycle condition}) is quotiented with
the equivalence relation of Eq.~(\ref{eq: coboundary equivalence}).
This equivalences class is an element of $H^2(G,U(1))$,
the second cohomology group of the group cohomology of $G$ over U(1).
Apparently, when phase functions of two states belong to 
different elements of $H^2(G,U(1))$,
we cannot adiabatically deform one state to the other
while preserving the symmetry.
Thus the cohomology group $H^2(G,U(1))$ classifies topological phases 
protected by symmetry
group $G$.\cite{fidkowski-kitaev11,chen-gu-wen11,chen-science12}
The definition of group cohomology and a useful formula
(K\"{u}nneth formula) in the calculation of non-trivial cocycles
are briefly summarized in Appendix~\ref{app: group cohomology}.

In an SPT phase characterized by a projective representation $U_g$
of symmetry group $G$,
the ground-state wave function possesses non-trivial boundary
modes of which symmetry transformations become anomalous.
To see this, we consider an MPS wave function on a finite chain of
length $L$,
\begin{align}
\ket{\Psi}= \sum_{\{m_i\}} v^\dagger
A^{m_{1}} A^{m_{2}} \dots A^{m_{L}} v'
\ket{m_{1} \ldots m_{L}},
\end{align}
where $v$ and $v'$ are boundary vectors specifying boundary
conditions at the end sites $1$ and $L$.
From Eq.~(\ref{eq: symmetry transformation of MPS}),
the action of an element $g$ of symmetry group $G$
transforms the MPS wave function as
\begin{align}
g\ket{\Psi}=e^{iL\theta_g} \sum_{\{m_i\}} v^\dagger
U_g^{-1} A^{m_{1}} A^{m_{2}} \dots &A^{m_{L}} U_g v'
 \ket{m_{1} \ldots m_{L}}.
\label{eq: transformation of boundary states}
\end{align}
Thus the boundary states determined by $v$ and $v'$ are transformed
according to $U_g$.
This indicates that the symmetry operations for effective boundary states
are not given by the original action of $g$ but by
its projective representation $U_g$.
In this sense the symmetry actions become anomalous at the boundaries.

\subsection{Haldane phase and SU(2) AKLT model}
The Haldane phase of $S=1$ antiferromagnetic spin chains is known
as an example of an SPT phase with symmetry
group $G=Z_2 \times Z_2$.\cite{pollmann10,pollmann11}
Let us consider the AKLT model,\cite{AKLT-prl87,AKLT-cmp88}
of which Hamiltonian reads
\begin{align}
H_\mathrm{AKLT}=\sum_i\left[
\bm{S}_i \cdot \bm{S}_{i+1} + \frac{1}{3}(\bm{S}_i \cdot \bm{S}_{i+1})^2
\right],
\label{eq: SU(2) AKLT Hamiltonian}
\end{align}
where $\bm{S}_i$ is a spin operator of $S=1$:
\begin{subequations}
\label{S=1 spin}
\begin{align}
S^x &=\frac{1}{\sqrt 2}
\begin{pmatrix}
 & 1 & \\
1 & & 1 \\
 & 1 & \\
\end{pmatrix}, \\
S^y &=\frac{1}{\sqrt 2}
\begin{pmatrix}
 & -i & \\
i & & -i \\
 & i & \\
\end{pmatrix}, \\
S^z &=
\begin{pmatrix}
1 &  & \\
 & 0 &  \\
 &  & -1 \\
\end{pmatrix}.
\label{S=1 Sz}
\end{align}
\end{subequations}
The AKLT Hamiltonian has SU(2) symmetry generated by
the above three spin operators.
In particular, they are invariant under its subgroup $Z_2 \times Z_2$
generated by a $\pi$-rotation around the $x$-axis,
\begin{align}
C_x = e^{i\pi S_x} =
\begin{pmatrix}
&& -1 \\
& -1 & \\
-1 && \\
\end{pmatrix},
\end{align}
and a $\pi$-rotation around the $z$-axis,
\begin{align}
C_z = e^{i\pi S_z} =
\begin{pmatrix}
-1&& \\
& 1 & \\
 && -1 \\
\end{pmatrix},
\end{align}
that commute with each other,
\begin{align}
C_x C_z=C_z C_x.
\end{align}

The ground state of the AKLT Hamiltonian is best described
in terms of the MPS in the following way.
We first decompose every $S=1$ spin into two $S=1/2$ spins.
Then the ground state is given as a valence-bond solid state
of virtual $S=1/2$ spins.
Namely, the ground-state wave function is obtained
by (i) projecting two $S=1/2$ spins from two neighboring sites
into a singlet state ($S=0$) and (ii) projecting the $S=1/2$ spins on each site
into a triplet state ($S=1$).
This is expressed in the MPS with 2 by 2 matrices
acting on the two-dimensional Hilbert space
of a virtual $S=1/2$ spin spanned by
$\{ \ket{\!\uparrow\,},\ket{\!\downarrow\,} \}$.
Two $S=1/2$ spins forming an $S=1$ spin on one site are coupled through
three types of matrices which are
the projection operators onto triplet states 
and labeled by the values of the total $S^z$:
\begin{align}
\tilde A^1&=
\begin{pmatrix}
1 & 0\\
0 & 0 \\
\end{pmatrix}, &
\tilde A^0&=
\frac{1}{\sqrt{2}}
\begin{pmatrix}
0 & 1 \\
1 & 0 \\
\end{pmatrix}, &
\tilde A^{-1}&=
\begin{pmatrix}
0 & 0 \\
0 & 1 \\
\end{pmatrix}.
\end{align}
Two $S=1/2$ spins from neighboring sites
are coupled by the matrix which is a projection operator
to a singlet state,
\begin{align}
\tilde B=
\frac{1}{\sqrt{2}}
\begin{pmatrix}
0 & 1 \\
-1 & 0 \\
\end{pmatrix}.
\end{align}
The ground-state wave function $\ket{\Psi}$ is then written
in the MPS form,
\begin{align}
\ket{\Psi}&=\sum_{\{m_i=0,\pm1\}}
\mathrm{tr}\left[A^{m_{1}} A^{m_{2}} \dots A^{m_{L}}\right]
\ket{m_{1} \ldots m_{L}},
\label{eq: MPS SU(2) AKLT}
\end{align}
where the matrices $A^m=\tilde A^m \tilde B$ are given by
\begin{align}
A^{\pm 1}&= \frac{\pm \sigma_x +i \sigma_y}{2 \sqrt{2}},&
A^0&= -\frac{1}{2}\sigma_z
\end{align}
in terms of the Pauli matrices $\sigma_{x,y,z}$.
This construction from projection operators is natural
because the AKLT Hamiltonian in Eq.~(\ref{eq: SU(2) AKLT Hamiltonian})
consists of a product of Casimir operators of neighboring $S=1$ spins
that project 
$\bm 3 \tensor \bm 3= \bm 5 \oplus \bm 3 \oplus \bm 1$
states onto $\bm 5$ states
such that the ground state is made of either $\bm 1$ or $\bm 3$ states
of neighboring spins,
i.e.,
two out of four $S=1/2$ spins on two neighboring sites form a singlet.

Now let us discuss transformation of the MPS $\ket{\Psi}$ by operators
from the $Z_2 \times Z_2$ symmetry group, i.e., $C_x, C_z$, and $C_xC_z$.
We can easily check that
\begin{align}
\sum_n (C_x)_{mn} A^n &= \sigma_x A^m \sigma_x, \n
\sum_n (C_z)_{mn} A^n &= \sigma_z A^m \sigma_z, \\
\sum_n (C_xC_z)_{mn} A^n &= \sigma_y A^m \sigma_y. \nonumber
\end{align}
Comparing these equations with Eq.\ (\ref{eq: symmetry transformation of MPS}),
we find a projective representation of the symmetry group
\begin{align}
(U_{C_x},U_{C_z},U_{C_x C_z})=(\sigma_x,\sigma_z,i\sigma_y),
\end{align}
and the associated phase function,
\begin{align}
\begin{aligned}
\phi(C_x, C_z)&=\pi, &
\phi(C_z, C_x C_z)&=0, &
\phi(C_x C_z, C_x)&=0, \\
\phi(C_z, C_x)&=0, &
\phi(C_x C_z, C_z)&=\pi, &
\phi(C_x, C_x C_z)&=\pi,
\end{aligned}
\end{align}
which is a 2-cocycle corresponding to a nontrivial element $\varphi$ of
$H^2(Z_2 \times Z_2, U(1))=\mathbb{Z}_2$ given in Appendix \ref{app: Z_N * Z_N}
 [Eq.~(\ref{eq: 2 cocycle ZN times ZN})].
We note that commuting operations $C_x$ and $C_z$
are represented projectively,
and their projective representations $\sigma_x$ and $\sigma_z$ anticommute with each other.

\section{$\mathbb{Z}_3$ SPT phase and SU(3) AKLT model \label{sec: SU(3) AKLT model}}

In this section we study 1D SPT phases which are protected by global
$Z_3 \times Z_3$ symmetry and characterized by a $\mathbb{Z}_3$
topological number.
They are natural generalizations of the
Haldane phase with $Z_2\times Z_2$ symmetry discussed in
the previous section.\cite{KD-quella-1-13,KD-quella-2-12,KD-quella-3-13}
We show that $\mathbb{Z}_3$ SPT phases are realized in
an SU(3) extension of the AKLT model.\cite{greiter-su3-07,greiter2-07}

\subsection{Group cohomology of $G=Z_3 \times Z_3$}
Here we present
a projective representation $U_g$ for
the symmetry group $G=Z_N \times Z_N$,
summarizing the results from Appendix~\ref{app: Z_N * Z_N}.

The group elements of $G$ are given by
Eq.\ (\ref{eq: group element of Z_N * Z_N}), and
its second cohomology group is $H^2(G,U(1))=\mathbb{Z}_N$,
generated by a 2-cocycle $\varphi$
shown in Eq.\ (\ref{eq: 2 cocycle ZN times ZN}).
A projective representation of $G$
with the phase function $\phi=\varphi$
is generated by $N\times N$ matrices
\begin{align}
U_x &=
\begin{pmatrix}
 & 1 &  &  \\
 &   & \ddots &  \\
 &   &  & 1  \\
1 &  &  &  \\
\end{pmatrix},
\qquad
U_y =
\begin{pmatrix}
1 &  &  &  \\
 & \omega^{}_N &  & \\
 &  & \ddots  &\\
 &  &  & \omega^{N-1}_N\\
\end{pmatrix},
\end{align}
which satisfy the algebra
\begin{subequations}
\begin{align}
U_x^N=U_y^N=1_N,
\qquad
U_x U_y = \omega^{}_N U_y U_x
\end{align}
with
\begin{equation}
\omega_N^{} = e^{2\pi i/N}.
\end{equation}
\end{subequations}

In the case of our main interest, $N=3$,
the projective representation is given by $3\times3$ matrices,
\begin{align}
U_x &=
\begin{pmatrix}
0 & 1 & 0 \\
0 & 0 & 1 \\
1 & 0 & 0 \\
\end{pmatrix},
\qquad
U_y =
\begin{pmatrix}
1 & 0 & 0 \\
0 & \omega & 0 \\
0 & 0 & \omega^2 \\
\end{pmatrix},
\label{eq: cocycle Z3*Z3}
\end{align}
where $\omega=\omega_3=\exp(2\pi i/3)$.
We thus expect a ground-state wave function of a $\mathbb{Z}_3$ SPT
phase to have the MPS form of $3\times3$ matrices which are
subject to symmetry transformations generated by
$U_x$ and $U_y$ in Eq.\ (\ref{eq: cocycle Z3*Z3}).
We will demonstrate this below.

\subsection{SU(3) AKLT model\label{sec: SU(3) AKLT model B}}

\begin{figure}
\begin{center}
\includegraphics[width=\linewidth]{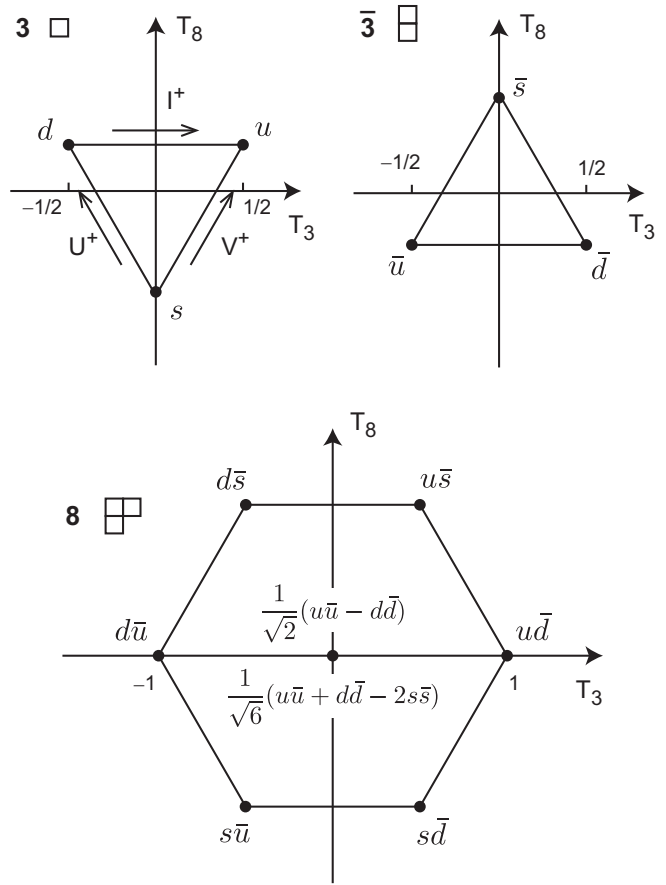}
\end{center}
\caption{
Weight diagrams of the Lie algebra su(3).
Fundamental representation $\bm 3$ consists of three
basis states $u,d,s$ (quarks).
Its conjugate representation $\bm\bar{\bm 3}$ consists of three
basis states $\bar u, \bar d, \bar s$ (antiquarks).
Adjoint representation $\bm 8$ is spanned by eight bilinear forms of
quarks and antiquarks, i.e., mesons.
}
\label{Fig: weight diagram}
\end{figure}

In this section we show that
an SPT phase protected by global $Z_3 \times Z_3$ symmetry
is realized in an SU(3) extension of the AKLT model.
We begin with a brief review on
representations of the Lie algebra su(3).\cite{geourgi-Lie-algebras}
In this paper, the three basis states
of the fundamental representation $\bm 3$ of su(3)
are denoted by three quarks $u,d,s$.
Similarly, its conjugate representation $\bm\bar{\bm 3}$ is spanned
by antiquarks $\bar u, \bar d, \bar s$.
We write the eight generators of su(3) in each representation as
$T^a$ ($a=1,\ldots,8$).
For the fundamental representation,
the su(3) generators are given by
\begin{align}
T^a=\frac{1}{2}\lambda_a,
\label{T^a in 3}
\end{align}
where $\lambda_a$'s are the Gell-Mann matrices:
\begin{align}
\label{Gell-Mann matrices}
\begin{aligned}
\lambda_1 &=
\begin{pmatrix}
0 & 1 & 0 \\
1 & 0 & 0 \\
0 & 0 & 0 \\
\end{pmatrix}, &
\lambda_2 &=
\begin{pmatrix}
0 & -i & 0 \\
i & 0 & 0 \\
0 & 0 & 0 \\
\end{pmatrix}, \\
\lambda_3 &=
\begin{pmatrix}
1 & 0 & 0 \\
0 & -1 & 0 \\
0 & 0 & 0 \\
\end{pmatrix}, &
\lambda_4 &=
\begin{pmatrix}
0 & 0 & 1 \\
0 & 0 & 0 \\
1 & 0 & 0 \\
\end{pmatrix}, \\
\lambda_5 &=
\begin{pmatrix}
0 & 0 & -i \\
0 & 0 & 0 \\
i & 0 & 0 \\
\end{pmatrix}, &
\lambda_6 &=
\begin{pmatrix}
0 & 0 & 0 \\
0 & 0 & 1 \\
0 & 1 & 0 \\
\end{pmatrix}, \\
\lambda_7 &=
\begin{pmatrix}
0 & 0 & 0 \\
0 & 0 & -i \\
0 & i & 0 \\
\end{pmatrix}, &
\lambda_8 &=
\frac{1}{\sqrt{3}}
\begin{pmatrix}
1 & 0 & 0 \\
0 & 1 & 0 \\
0 & 0 & -2 \\
\end{pmatrix}.
\end{aligned}
\end{align}
For the conjugate representation $\bm\bar{\bm 3}$,
the su(3) generators are given by
\begin{align}
T^a=-\frac{1}{2}\lambda^*_a.
\end{align}
Cartan subalgebra of su(3) consists of $T^3$ and $T^8$
that allow us to define weight vectors.
The weight diagrams of the fundamental representation $\bm 3$
and its conjugate representation $\bm\bar{\bm 3}$
are shown in the $T^3$-$T^8$ plane in Fig.~\ref{Fig: weight diagram}.
The raising and lowering operators defined by
\begin{align}
I^\pm&\equiv T^1\pm i T^2,&
U^\pm&\equiv T^6\pm i T^7,&
V^\pm&\equiv T^4\pm i T^5,
\label{eq: raising and lowering ops}
\end{align}
are also indicated for the fundamental representation $\bm 3$
in Fig.~\ref{Fig: weight diagram}.

\begin{figure}[tb]
\begin{center}
\includegraphics[width=\linewidth]{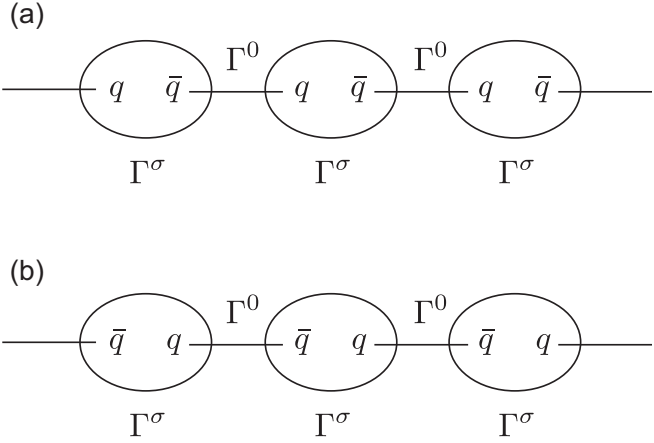}
\end{center}
\caption{
Schematic pictures of the two types of MPS wave functions for
the SU(3) AKLT model.
On every site there are $\bm 3$ states (quark $q$)
and $\bm\bar{\bm 3}$ states (antiquark $\bar q$)
which are projected onto $\bm 8$ states (mesons represented by ovals)
through the eight traceless matrices $\Gamma^\sigma$.
(a) $q$ is coupled through the $\Gamma^0$ matrix
to $\bar q$ on the left neighboring site
to form a singlet $\bm1$ ($\eta'$ meson).
(b) $q$ is coupled through the $\Gamma^0$ matrix
to $\bar q$ on the right neighboring site
to form a singlet $\bm1$ ($\eta'$ meson).
}
\label{Fig: SU(3) AKLT wave function}
\end{figure}

The SU(3) extension of the AKLT state
is obtained as follows.
We assume that both $\bm 3$ representation ($u$, $d$, $s$)
and $\bm\bar{\bm 3}$ representation ($\bar u$, $\bar d$, $\bar s$)
are placed on each site.
From their tensor product,
\begin{align}
\bm 3 \tensor \bm\bar{\bm 3} = \bm 8 \oplus \bm 1,
\label{3 & 3^*}
\end{align}
we keep the octet representation $\bm 8$ on each site.
This is analogous to keeping an on-site triplet in the SU(2) case.
For each pair of neighboring sites, we combine $\bm 3$ from one site
and $\bm\bar{\bm 3}$ from the other and project them onto singlet $\bm 1$,
again similarly to the SU(2) case.
At each end of a finite open chain,
we have unpaired $\bm 3$ or $\bm\bar{\bm 3}$ states,
which form a triplet zero-energy boundary mode.
Figure~\ref{Fig: SU(3) AKLT wave function} shows
schematic pictures of the SU(3) AKLT states.
We note that there are two ways of constructing such states;
see Fig.~\ref{Fig: SU(3) AKLT wave function}(a) and (b).\cite{greiter2-07}
Here we first discuss the state
shown in Fig.~\ref{Fig: SU(3) AKLT wave function}(a) in detail.
The other state will be discussed in Sec.~\ref{sec: two types}.

On each site we have $\bm 8$ states in the bilinear form
of $\bm 3$ and $\bm\bar{\bm 3}$ states coupled by
$3 \times 3$ traceless matrices $\Gamma^i~ (i=1,\ldots,8)$,
\begin{align}
\Vec{\Gamma}=(I^+, I^-, \sqrt{2}T^3, V^-, V^+, U^+, U^-, \sqrt{2}T^8)^T,
\label{eq: octet SU(3)}
\end{align}
where the su(3) operators $T^a$ are the ones defined
in Eq.\ (\ref{T^a in 3}) and the raising and lowering operators are
defined in Eq.\ (\ref{eq: raising and lowering ops}).
The eight states $\ket{\bm \sigma}$ form the adjoint representation
$\bm 8$ of su(3),
corresponding to the octet of mesons:
\begin{align}
\ket{\bm \sigma}
&=(\ket{u}, \ket{d}, \ket{s})
\Vec{\Gamma}
\begin{pmatrix}
\ket{\bar u} \\
\ket{\bar d} \\
\ket{\bar s} \\
\end{pmatrix} \n
&=
\begin{pmatrix}
\pi^+ \\ \pi^- \\ \pi^0 \\ K^- \\ K^+ \\ K^0 \\ \bar K^0 \\ \eta
\end{pmatrix}
=
\begin{pmatrix}
\ket{u \bar d} \\ \ket{d \bar u} \\ \frac{1}{\sqrt2}\ket{u \bar u- d \bar d} \\
\ket{s \bar u} \\ \ket{u\bar s} \\ \ket{d \bar s}  \\ \ket{s \bar d} \\
\frac{1}{\sqrt 6} \ket{u \bar u + d \bar d -2s \bar s}
\end{pmatrix},
\label{8 mesons}
\end{align}
whose weight diagram is shown in Fig.~\ref{Fig: weight diagram}.
The su(3) generators $T^a$ for the $\bm 8$ representation
are given by $8\times 8$ matrices,
which are written in this basis as
\begin{align}
(T^a)_{ij}=
\frac{1}{2}\t{tr}[(\Gamma^i)^\dagger (\lambda_a \Gamma^j - \Gamma^j \lambda_a)].
\label{T^a in adjoint rep}
\end{align}
With a basis transformation
[replacing $\Gamma^i$ with $\lambda_i/\sqrt{2}$ in Eqs.~(\ref{8 mesons})
and (\ref{T^a in adjoint rep})],
we can rewrite $T^a$ in the standard form
\begin{align}
(T^a)_{ij}= -i f_{aij},
\end{align}
where $f_{aij}$ is the structure constant of su(3)
defined from the commutation relation
$
[T^a, T^b]=if_{abc}T^c.
$

The singlet state $\bm{1}$ on each bond is given by
the bilinear form
\begin{align}
\eta'=\frac{1}{\sqrt{3}}\ket{u \bar u + d \bar d + s \bar s},
\end{align}
which is composed of $\bm 3$ and $\bm{\bar{3}}$ states from
neighboring sites coupled through the matrix
\begin{align}
(\Gamma^0)_{ij}=\frac{1}{\sqrt{3}}\delta_{ij}.
\end{align}

The MPS wave function of the SU(3) AKLT state shown in
Fig.~\ref{Fig: SU(3) AKLT wave function}(a) is constructed as follows.
First, $\bm 3$ and $\bm\bar{\bm 3}$ states from neighboring sites
are projected onto the singlet $\bm 1$ using the $\Gamma^0$ matrix,
\begin{align}
|\psi_{\bm 1} \rangle &=
\sum_{\{a_i\}} \sum_{\{b_i\}}
\Gamma^0_{b_1 a_2} \Gamma^0_{b_2 a_3} \ldots
\Gamma^0_{b_{L-1} a_L} \Gamma^0_{b_L a_1}
| a_1 b_1 \ldots a_L b_L \rangle ,
\label{MPS Gamma^0}
\end{align}
where $a_i$ and $b_i$ label states of
$\bm 3$ representation and $\bm\bar{\bm 3}$ representation, respectively,
on site $i$ in a 1D periodic chain of length $L$.
Second, $\bm 3$ and $\bm\bar{\bm 3}$ states on the same site
are projected onto $\bm 8$ states
using the eight traceless matrices $\Gamma^\sigma$
in Eq.\ (\ref{eq: octet SU(3)}),
\begin{align}
|\psi_\t{sym} \rangle &=
\sum_{\{\sigma_i\}} \sum_{\{a_i\}} \sum_{\{b_i\}}
\Gamma^{\sigma_1}_{a_1 b_1} \Gamma^0_{b_1 a_2}
 \Gamma^{\sigma_2}_{a_2 b_2} \Gamma^0_{b_2 a_3} \ldots  \n
&\qquad\times \Gamma^{\sigma_{L-1}}_{a_{L-1} b_{L-1}}
 \Gamma^0_{b_{L-1} a_L} \Gamma^{\sigma_L}_{a_L b_L}
 \Gamma^0_{b_L a_1} | \sigma_1 \ldots \sigma_L \rangle ,
\label{MPS Gamma^sigma Gamma^0}
\end{align}
where $\sigma_i$ labels physical states in the $\bm 8$ representation
of Eq.\ (\ref{8 mesons}).
Finally, the SU(3) AKLT wave function is obtained as
\begin{subequations}
\label{eq: MPS SU(3) AKLT}
\begin{align}
|\Psi \rangle &=
 C^{-1} \sum_\sigma \t{tr}[A^{\sigma_1} A^{\sigma_2} \ldots A^{\sigma_L}]
| \sigma_1\ldots\sigma_L \rangle,
\end{align}
with
\begin{align}
A^{\sigma_i} &=\Gamma^{\sigma_i}\Gamma^0
= \frac{1}{\sqrt{3}} \Gamma^{\sigma_i}.
\end{align}
The normalization constant is
\begin{align}
C&=\langle \psi_\t{sym}|\psi_\t{sym} \rangle
=\left(\frac{8}{9}\right)^{L/2} .
\end{align}
\end{subequations}
The wave function $\ket{\Psi}$ 
has the same MPS form as Eq.\ (\ref{eq: MPS periodic})
up to the normalization factor.
Alternatively,
we can write the MPS wave function $\ket{\Psi}$ as\cite{greiter2-07}
\begin{subequations}
\label{eq: MPS2 SU(3) AKLT}
\begin{align}
|\Psi\rangle&=\frac{1}{3^{L/2} C}{\rm tr} [M_1 M_2\cdots M_L]
\label{eq: MPS2 SU(3) AKLT a}
\end{align}
with matrices taking values in the local Hilbert space,
\begin{widetext}
\begin{align}
M_i&=\left(
  \begin{array}{ccc}
\frac23|u\bar{u}\rangle_i-\frac13 |d\bar{d}\rangle_i-\frac13 |s\bar{s}\rangle_i
  & |u\bar{d}\rangle_i & |u\bar{s}\rangle_i \\
|d\bar{u}\rangle_i &
-\frac13|u\bar{u}\rangle_i+\frac23 |d\bar{d}\rangle_i-\frac13 |s\bar{s}\rangle_i
  & |d\bar{s}\rangle_i \\
|s\bar{u}\rangle_i & |s\bar{d}\rangle_i
  & -\frac13 |u\bar{u}\rangle_i-\frac13 |d\bar{d}\rangle_i
    +\frac23 |s\bar{s}\rangle_i
  \end{array}
\right).
\label{M_i}
\end{align}
\end{widetext}
\end{subequations}

Let us construct a Hamiltonian having the above MPS wave function
as a ground state, using projection operators acting on the
$\bm 8$ representations (mesons) on every pair of neighboring states,
in the same way as in the SU(2) AKLT model.
The product of two sets of $\bm 8$ states
from neighboring sites
is decomposed as
\begin{align}
\bm 8 \tensor \bm 8 =
\bm{27} \oplus \bm{10} \oplus \overbar{\bm{10}} \oplus \bm 8
 \oplus \bm 8 \oplus \bm 1.
\label{eq: decomposition 8 times 8}
\end{align}
However, the formation of a singlet on every bond,
which was imposed in Eq.\ (\ref{MPS Gamma^0}),
implies that the maximum multiplets that can be formed
by states from two neighboring sites 
are actually limited to
\begin{align}
\bm 3 \tensor \bm\bar{\bm 3} = \bm 8 \oplus \bm 1.
\end{align}
Therefore, if a Hamiltonian is a projection operator
annihilating both $\bm 8$ and $\bm 1$ representations 
for every pair of $\bm 8$ states of neighboring sites
in Eq.~(\ref{eq: decomposition 8 times 8}),
then the MPS wave function $\ket{\Psi}$ in Eq.\ (\ref{eq: MPS SU(3) AKLT})
becomes a zero-energy eigenstate.
We can write down such a Hamiltonian using Casimir operators as
\begin{align}
H_3&=\frac{1}{4}\sum_{i}
[({\bm T}_i + {\bm T}_{i+1})^2 - C(\bm 8)]
[({\bm T}_i + {\bm T}_{i+1})^2 - C(\bm 1)],
\end{align}
where $T^a$'s are su(3) operators in the $\bm 8$ representation
given in Eq.\ (\ref{T^a in adjoint rep}),
and $C(\bm{d})$ is the eigenvalue of the quadratic Casimir operator,
$\sum_a (T^a)^2$,
for $d$-dimensional representations.
The MPS wave function in Eq.~(\ref{eq: MPS SU(3) AKLT}) is
a zero-energy eigenstate of $H_3$,
whose eigenvalues are non-negative by construction.
Hence the MPS state is an exact ground state.
The other SU(3) AKLT state shown in
Fig.~\ref{Fig: SU(3) AKLT wave function}(b) is another
zero-energy ground state of $H_3$, and there is a finite
energy gap to excited states.\cite{greiter2-07}
Using
\begin{align}
C(\bm 8)&=3, &
C(\bm 3)&=C(\bm\bar{\bm{3}})=\frac{4}{3}, &
C(\bm 1)&=0,
\end{align}
we can reduce the Hamiltonian to the simpler form
\begin{align}
H_3&=\sum_{i}\left[
({\bm T}_i \cdot {\bm T}_{i+1})^2
 + \frac{9}{2} {\bm T}_i \cdot {\bm T}_{i+1}
 + \frac{9}{2}
\right],
\label{eq: SU(3) AKLT Hamiltonian}
\end{align}
which we shall call the SU(3) AKLT Hamiltonian
in the rest of this paper.
We note once again that the su(3) generators $\bm{T}_i$ are
in the $\bm8$ representation.

\subsection{Symmetry operations of $Z_3\times Z_3$}

Here we derive symmetry actions of the $Z_3 \times Z_3$ symmetry 
on the eight physical states $\ket{\bm\sigma}$
by using the projective representation $U_{x,y}$
[Eq.~(\ref{eq: cocycle Z3*Z3})]
and Eq.~(\ref{eq: symmetry transformation of MPS}).
The operations of $U_x$ and $U_y$ on the octet of matrices
$\Gamma^1, \ldots, \Gamma^8$
yield
\begin{subequations}
\begin{equation}
\begin{aligned}
&
U_x^{-1} (\Gamma^1, \Gamma^4, \Gamma^6) U_x = (\Gamma^6, \Gamma^1, \Gamma^4), \\
&
U_x^{-1} (\Gamma^2, \Gamma^5, \Gamma^7) U_x = (\Gamma^7, \Gamma^2, \Gamma^5), \\
&
U_x^{-1} (\Gamma^3, \Gamma^8) U_x =
\frac{1}{2}(-\Gamma^3+\sqrt{3}\Gamma^8,-\sqrt{3}\Gamma^3-\Gamma^8),
\end{aligned}
\end{equation}
and
\begin{align}
\begin{aligned}
&
U_y^{-1} (\Gamma^1, \Gamma^4, \Gamma^6) U_y
 = \omega(\Gamma^1, \Gamma^4, \Gamma^6), \\
&
U_y^{-1} (\Gamma^2, \Gamma^5, \Gamma^7) U_y
 = \omega^2(\Gamma^2, \Gamma^5, \Gamma^7), \\
&
U_y^{-1} (\Gamma^3, \Gamma^8) U_y
 = (\Gamma^3, \Gamma^8 ).
\end{aligned}
\end{align}%
\label{eq: Z3 * Z3 for mesons}%
\end{subequations}
These relations determine actions of the $Z_3\times Z_3$ symmetry
operators on the eight matrices $\Gamma_1,\ldots,\Gamma_8$,
which we write in the form of
Eq.~(\ref{eq: symmetry transformation of MPS}) as
\begin{equation}
\sum_{n=1}^8 x_{mn}\Gamma^n=U_x^{-1}\Gamma^m U_x,
\qquad
\sum_{n=1}^8 y_{mn}\Gamma^n=U_y^{-1}\Gamma^m U_y,
\label{x Gamma, y Gamma}
\end{equation}
where
\begin{subequations}
\label{x and y}
\begin{align}
x&=
\begin{pmatrix}
 0 & 0 & 0 & 0 & 0 & 1 & 0 & 0 \\
 0 & 0 & 0 & 0 & 0 & 0 & 1 & 0 \\
 0 & 0 & -\frac{1}{2} & 0 & 0 & 0 & 0 &
   \frac{\sqrt{3}}{2} \\
 1 & 0 & 0 & 0 & 0 & 0 & 0 & 0 \\
 0 & 1 & 0 & 0 & 0 & 0 & 0 & 0 \\
 0 & 0 & 0 & 1 & 0 & 0 & 0 & 0 \\
 0 & 0 & 0 & 0 & 1 & 0 & 0 & 0 \\
 0 & 0 & -\frac{\sqrt{3}}{2} & 0 & 0 & 0 & 0 &
   -\frac{1}{2} \\
\end{pmatrix}, \\
y &=
\begin{pmatrix}
 \omega & 0 & 0 & 0  & 0 & 0 & 0 & 0 \\
 0 & \omega^2 & 0 & 0  & 0 & 0 & 0 & 0 \\
 0 & 0 & 1 & 0 & 0 & 0 & 0 & 0 \\
 0 & 0 & 0 & \omega & 0 & 0 & 0 & 0 \\
 0 & 0 & 0 & 0 & \omega^2 & 0 & 0 & 0 \\
 0 & 0 & 0 & 0 & 0 & \omega & 0 & 0 \\
 0 & 0 & 0 & 0 & 0 & 0 & \omega^2 & 0 \\
 0 & 0 & 0 & 0 & 0 & 0 & 0 & 1 \\
\end{pmatrix}.
\end{align}
\end{subequations}
The $8\times8$ matrices $x$ and $y$ are generators of an
eight-dimensional representation of
the symmetry group $G=Z_3 \times Z_3$, satisfying
\begin{equation}
x^3=y^3=1, \qquad [x,y]=0.
\end{equation}
By contrast, $U_x$ and $U_y$ in Eq.\ (\ref{eq: cocycle Z3*Z3})
make a projective representation of the symmetry group $G$
with the phase function $\phi=\varphi$ defined in
Eq.\ (\ref{eq: 2 cocycle ZN times ZN}),
which indicates that the ground state $\ket{\Psi}$
is in the SPT phase of $1\in\mathbb{Z}_3=\{0,1,2\}$.
The Hamiltonian in Eq.\ (\ref{eq: SU(3) AKLT Hamiltonian}),
which is made of Casimir operators of su(3),
is invariant under SU(3) and therefore invariant under
the symmetry group $G=Z_3\times Z_3$,
a subgroup of SU(3).
More importantly, the MPS wave function (\ref{eq: MPS SU(3) AKLT})
is also invariant under $G=Z_3\times Z_3$ because
the eight constituent matrices, $A^\sigma=\Gamma^\sigma/\sqrt3$,
satisfy the transformation relations (\ref{x Gamma, y Gamma}).

\subsection{Two types of MPSs and two $\mathbb{Z}_3$ SPT phases\label{sec: two types}}
Let us consider the other ground-state wave function of $H_3$,
i.e., the SU(3) AKLT state $\ket{\widetilde{\Psi}}$
shown in Fig.~\ref{Fig: SU(3) AKLT wave function}(b).
Comparing Figs.~\ref{Fig: SU(3) AKLT wave function}(a) and (b),
we see that $\ket{\Psi}$ and $\ket{\widetilde{\Psi}}$ are related to
each other by spatial inversion.\cite{greiter2-07}
From Eq.~(\ref{eq: MPS SU(3) AKLT})
we can write the MPS representation of $\ket{\widetilde{\Psi}}$ as
\begin{subequations}
\label{eq: MPS tilde Psi}
\begin{align}
|\widetilde{\Psi} \rangle &=
 C^{-1} \sum_\sigma \t{tr}\!\left(
\widetilde{A}^{\sigma_1} \widetilde{A}^{\sigma_2}
 \ldots \widetilde{A}^{\sigma_L}
\right)
| \sigma_1\ldots\sigma_L \rangle,
\end{align}
where
\begin{align}
\widetilde{A}^\sigma &= (A^\sigma)^T.
\end{align}
\end{subequations}
Similarly, following Eq.~(\ref{eq: MPS2 SU(3) AKLT}),
we can rewrite $\ket{\widetilde{\Psi}}$ as
\begin{equation}
\ket{\widetilde{\Psi}}=
\frac{1}{3^{L/2}C}
\mathrm{tr}\!\left(\widetilde{M}_1\widetilde{M}_2\cdots\widetilde{M}_L\right)
\end{equation}
with $\widetilde{M}=M^T$.

The two states $\ket{\Psi}$ and $\ket{\widetilde{\Psi}}$
are orthogonal in the thermodynamic limit.
Actually, the overlap of the states vanishes,
\begin{align}
\langle\Psi|\widetilde\Psi\rangle 
\propto
(\epsilon'/\epsilon)^L,
\end{align}
as $L\to \infinity$,
where $\epsilon$ and $\epsilon'$ are
the largest eigenvalues in magnitude of transfer matrices
$\mathcal{M}=\sum_m A^m \tensor (A^m)^*$
and 
$\mathcal{M}'=\sum_m (A^m)^T \tensor (A^m)^*$,
respectively
($\epsilon=8/9, \epsilon'=-4/9$).

The MPS wave function $\ket{\widetilde{\Psi}}$
describes a ground state in one of the $\mathbb{Z}_3$
SPT phases (i.e., $2\in \mathbb{Z}_3$)
and realizes a projective representation of 
the $Z_3 \times Z_3$ symmetry as follows.
The operations of the generators $x,y$ of the $Z_3 \times Z_3$ symmetry
in Eq.~(\ref{x and y})
on the wave function $\ket{\widetilde{\Psi}}$
induces the transformations of the matrices $\widetilde{A}^n$
\begin{subequations}
\begin{align}
\sum_{n=1}^8 x_{mn}\widetilde{A}^n =
\widetilde{U}_x^{-1} \widetilde{A}^m \widetilde{U}_x,
\qquad
\sum_{n=1}^8 y_{mn}\widetilde{A}^n =
\widetilde{U}_y^{-1} \widetilde{A}^m \widetilde{U}_y,
\end{align}
with
\begin{equation}
\widetilde{U}_x =
\begin{pmatrix}
0&1&0\\
0&0&1\\
1&0&0\\
\end{pmatrix},
\qquad
\widetilde{U}_y =
\begin{pmatrix}
1&0&0\\
0&\omega^2&0\\
0&0&\omega\\
\end{pmatrix}.
\end{equation}
\end{subequations}
These two matrices $\widetilde{U}_x, \widetilde{U}_y$
give a projective representation of $Z_3\times Z_3$
and satisfy
\begin{align}
\widetilde{U}_x^3=\widetilde{U}_y^3=1_3,
\qquad
\widetilde{U}_x \widetilde{U}_y = \omega^2 \widetilde{U}_y \widetilde{U}_x.
\end{align}
The phase function $\tilde \phi$ in this projective representation
is a nontrivial 2-cocycle and given by
\begin{align}
\tilde \phi=2\varphi
\end{align}
with $\varphi$ in Eq.~(\ref{eq: 2 cocycle ZN times ZN}).
Thus the MPS wave function $\ket{\widetilde{\Psi}}$
belongs to the SPT phase of $2 \in \mathbb{Z}_3$.
To summarize, both MPS wave functions,
$\ket{\Psi}$ and $\ket{\widetilde{\Psi}}$ made of matrices
$A^n$ and $\widetilde{A}^n$ respectively,
are zero-energy ground states
of the SU(3) AKLT Hamiltonian [Eq.~(\ref{eq: SU(3) AKLT Hamiltonian})]
and belong to two different SPT phases
which are characterized by the $\mathbb{Z}_3$ topological
index as 1 and 2 ($\in\mathbb{Z}_3$),
respectively.

Actually, the two states $\ket{\Psi}$ and $\ket{\widetilde{\Psi}}$
should be considered as two-fold degenerate SPT ground ``states''
in a {\em single} gapped SPT phase.
The situation is similar to the 
two-dimensional ferromagnetic Ising model, 
where the low-temperature ordered phase is
a single gapped phase with two-fold degenerate ground states
with ferromagnetic long-range order.
The degeneracy is lifted by applying
a finite magnetic field along the Ising spin direction,
and changing the sign of the magnetic field
leads to a first-order phase transition between the two
ferromagnetically ordered states.
Our SU(3) AKLT Hamiltonian is symmetric under inversion and
is similar to the ferromagnetic Ising model without a field.
In analogy with the Ising model, we expect that
the SU(3) AKLT model should have a first-order phase transition
between the two SPT ``states'' $\ket{\Psi}$ and $\ket{\widetilde{\Psi}}$
when we change the sign of an inversion symmetry breaking term
added to the model.
Without such a term, we have a single gapped phase with doubly degenerate
ground states under the periodic boundary condition.
Under open boundary conditions,
the inversion symmetry is manifestly
broken by the appearance of different kinds of
boundary zero modes 
($\bm{3}$ or $\bar{\bm{3}}$ states)
at the left and right boundaries,
and the ground state is 18-fold degenerate ($2\cdot3\cdot3=18$)
if we neglect exponentially small coupling between
the left and right boundary modes.
In Sec.~VIA, we will present an $S=1$ spin chain with staggered
quadrupole couplings which breaks the inversion symmetry explicitly.
There we find a unique ground state (under periodic boundary conditions)
that is a nontrivial $\mathbb{Z}_3$ SPT phase.

Finally, we note that the SU(3) AKLT model can also be thought of
as a PSU(3) AKLT model realizing $\mathbb{Z}_3$ SPT states protected by
PSU(3) symmetry.\cite{KD-quella-2-12,KD-quella-1-13}
This is because the adjoint representation $\bm{8}$ is also a representation
of PSU(3) and the SU(3) AKLT Hamiltonian $H_3$ respects the PSU(3) symmetry.
Furthermore, $\bm{3}$ or $\bar{\bm{3}}$ states appearing at the ends of
a spin chain give projective representations of PSU(3) corresponding to
$1$ and $-1 \in H^2(\mbox{PSU(3),\, U(1)})=\mathbb{Z}_3$, respectively. 
Thus the two states $\ket{\Psi}$ and $\ket{\widetilde{\Psi}}$
represent two $\mathbb{Z}_3$ SPT states
protected by PSU(3) symmetry group.

\subsection{SU(N) AKLT Hamiltonian \label{subsec: SU(N) AKLT}}
In a similar way to the SU(3) case,
we can obtain an SU($N$) generalization of the AKLT Hamiltonian
as follows.
We consider a 1D chain in which
the local Hilbert space on each site is 
spanned by the $\bm{N^2-1}$ (adjoint) representation of su($N$).
Then the SU($N$) AKLT Hamiltonian is constructed from
projection operators for two neighboring sites:
\begin{align}
H_N=\frac{1}{4}\sum_{i}
&[({\bm T}_i + {\bm T}_{i+1})^2 - C(\bm{N^2-1})] \n
\times&[({\bm T}_i + {\bm T}_{i+1})^2 - C(\bm 1)], 
\end{align}
where $T^a$ ($a=1,\ldots,N^2-1$) are su($N$) operators
in the $\bm{N^2-1}$ representation,
and $C(\bm{d})$ is the eigenvalue of a quadratic Casimir operator
for $d$-dimensional representations.
With the eigenvalues of Casimir operators
\begin{align}
C(\bm{N^2-1})&=N, & C(\bm{N})&=\frac{N^2-1}{2N}, & C(\bm{1})&=0,
\end{align}
the SU($N$) extension of the AKLT Hamiltonian is reduced to
\begin{align}
H_N&=
\sum_{i}\left[
({\bm T}_i \cdot {\bm T}_{i+1})^2
 + \frac{3 N}{2} {\bm T}_i \cdot {\bm T}_{i+1}
 + \frac{N^2}{2}
\right],
\end{align}
which we call the SU($N$) AKLT model.
\footnote{
After completion of this manuscript,
we became aware that $H_N$ and $\xi_N$ were reported in
S.~Rachel, D.~Schuricht, B.~Scharfenberger, R.~Thomale, M.~Greiter, J. Phys.: Conf. Ser. \textbf{200} 022049 (2010).}
It is a natural generalization of the SU(2) $S=1$ AKLT model
and the SU(3) AKLT model,
in that the generators $\bm{T}_i$ are in the adjoint representation
as in these two models.
Since the tensor product of two adjoint representations $\bm{N^2-1}$
of su($N$) ($N\ge 4$) are decomposed as
\begin{align}
&(\bm{N^2-1})\tensor (\bm{N^2-1}) \n
&=\bm{\frac{1}{4}N^2(N+3)(N-1)} \oplus \bm{\frac{1}{4}(N^2-1)(N^2-4)} \n
&\quad \oplus \overbar{\bm{\frac{1}{4}(N^2-1)(N^2-4)}} \oplus 
\bm{\frac{1}{4}N^2(N+1)(N-3)}  \n
&\quad \oplus (\bm{N^2-1}) \oplus (\bm{N^2-1}) \oplus \bm{1},
\end{align}
the energy spectrum of the SU($N$) AKLT Hamiltonian $H_N$ is non-negative.

We can easily construct the zero-energy ground-state wave function
of $H_N$ 
as an SU($N$) extension of the AKLT wave function.
Suppose that each site of a 1D chain consists of
virtual degrees of freedom spanned by
the fundamental representation $\bm N$ and its conjugate representation
$\bm{\overbar{N}}$ of su($N$).
Using the decomposition
\begin{align}
\bm N \tensor \bm{\overbar{N}} = (\bm{N^2-1}) \oplus \bm 1,
\end{align}
we construct an SU($N$) AKLT wave function 
by projecting virtual $\bm N$ and $\bm{\overbar{N}}$ states onto 
physical $\bm{N^2-1}$ states at every site
and onto the singlet $\bm 1$ at every bond, as we have done for $N=3$
in Sec.\ \ref{sec: SU(3) AKLT model B}.
We can write the SU($N$) AKLT wave function
in the MPS form analogous to Eq.\ (\ref{eq: MPS SU(3) AKLT})
by replacing $\Vec \Gamma$ with the generators of
the fundamental representation of su($N$).
By construction, the above SU($N$) AKLT wave function
is a zero-energy ground state of the SU($N$) AKLT Hamiltonian $H_N$.
Since the fundamental representation $\bm{N}$ and its
conjugate representation $\bm{\overbar{N}}$ are different for $N>2$,
the ground state of $H_N$ is twofold degenerate (under periodic
boundary conditions) as in the $N=3$ case
($\ket{\Psi}$ and $\ket{\widetilde\Psi}$
schematically shown in Fig.~\ref{Fig: SU(3) AKLT wave function}).
These degenerate ground states realize two distinct phases
of $\mathbb{Z}_N$ SPT phases
($1,-1\in \mathbb{Z}_N$), as one can verify by determining
the action of the $Z_N \times Z_N$ symmetry in a similar way to
Eq.~(\ref{x and y}).
However, as we have discussed earlier,
we should consider these two states as two-fold
degenerate ground states in a {\em single} SPT phase.
Under open boundary conditions the ground state is
$2N^2$-fold degenerate.

Let us compute the correlation functions of
the SU($N$) operators $T^a$ (in the $\bm{N^2-1}$ representation)
for the SU($N$) AKLT states.
We use the following properties of the transfer matrix
\begin{align}
\mathcal{M}=\sum_{m=1}^{N^2-1} A^m \tensor (A^m)^*,
\end{align}
which are derived in Appendix~\ref{app: SU(N) AKLT}.
The transfer matrix has eigenvectors satisfying
\begin{subequations}
\begin{align}
\mathcal{M}\ket{v_0}&=\epsilon_1 \ket{v_0}, \\
\mathcal{M}\ket{v_m}&=\epsilon_2 \ket{v_m}, \qquad (m=1,\ldots, N^2-1), 
\end{align}
with the ratio of the eigenvalues
\begin{align}
\frac{\epsilon_2}{\epsilon_1}&= \frac{-1}{N^2-1}.
\end{align}
\label{eq: eigs of M}
\end{subequations}
In order to compute the correlation function of $T^a$,
we define another transfer matrix
\begin{align}
\widetilde{\mathcal{M}}^a =\sum_{m,n=1}^{N^2-1} T^a_{mn} A^n \tensor (A^m)^*.
\end{align}
Then the correlation function for the SU($N$) AKLT state $\ket{\Psi}$
is written as
\begin{align}
\bra{\Psi}T^a_i T^a_j \ket{\Psi}
&=
\frac{
\bra{v_0} \widetilde{\mathcal{M}}^a \mathcal{M}^{i-j-1}
          \widetilde{\mathcal{M}}^a \ket{v_0}
}{
\bra{v_0} \mathcal{M}^{i-j+1} \ket{v_0}
},
\end{align}
where we assume $i>j$.
We can show that 
the vector $\ket{v_0}$ and $\widetilde{\mathcal{M}}^a \ket{v_0}$
are orthogonal,
\begin{align}
\bra{v_0} \widetilde{\mathcal{M}}^a \ket{v_0}=0
\label{eq: orthogonality of v0 and Ma v0}
\end{align}
for any $a=1,\ldots,N^2-1$,
which implies that $\widetilde{\mathcal{M}}^a \ket{v_0}$
is an eigenvector of $\mathcal{M}$ with the eigenvalue $\epsilon_2$.
For the derivation of Eqs.~(\ref{eq: eigs of M})
and (\ref{eq: orthogonality of v0 and Ma v0}),
see Appendix~\ref{app: SU(N) AKLT}.
Now we can compute the correlation function as
\begin{align}
\bra{\Psi}T^a_i T^a_j \ket{\Psi}
&=
\frac{
\epsilon_2^{i-j-1} \bra{v_0} \widetilde{\mathcal{M}}^a \widetilde{\mathcal{M}}^a \ket{v_0}
}{
\epsilon_1^{i-j+1} 
}\n
&\propto \left(\frac{-1}{N^2-1}\right)^{i-j}.
\end{align}
Thus the correlation function decays exponentially 
with the correlation length\cite{Note1}
\begin{align}
\xi_N=\frac{1}{\ln (N^2-1)},
\end{align}
which indicates the existence of
a finite energy gap between the ground state and excited states
of the SU($N$) AKLT Hamiltonian $H_N$.

\section{String order and hidden $Z_3\times Z_3$ symmetry breaking \label{sec: string order}}

In this section we discuss hidden order in the ground state of
the SU(3) AKLT model.
A hidden order characterized by the string order
parameter\cite{denNijsRommelse} was
first found in the Haldane phase,
where the nonvanishing string order corresponds to
a hidden $Z_2\times Z_2$ symmetry breaking
in the system obtained after a nonunitary
transformation.\cite{KennedyTasaki,KennedyTasaki2}
A generalization of the string order was discussed recently for SPT phases
with $Z_N\times Z_N$
symmetry.\cite{KD-quella-2-12,KD-quella-3-13}
Here we demonstrate the existence of a hidden order in the ground state
of the SU(3) AKLT model,
i.e., in the matrix product state in Eq.\ (\ref{eq: MPS SU(3) AKLT}).
Throughout this section we assume the symmetry of the system to be SU(3),
rather than $Z_3\times Z_3$
that we have assumed in the preceding sections.
Accordingly, the string order parameters that we define below
are different from
those studied in Refs.~\onlinecite{KD-quella-2-12,KD-quella-3-13}
and give natural generalization of the conventional string order of
the SU(2) symmetric Haldane phase.
However, assuming full SU(3) symmetry
has the disadvantage of losing direct contact with the nonlocal unitary
transformation with which the hidden order can be related to
$Z_3\times Z_3$ symmetry breaking.

We study hidden order in the matrix product state $\ket{\Psi}$
in Eq.\ (\ref{eq: MPS2 SU(3) AKLT}).
Let us consider the following ``up'' operator
\begin{align}
{\cal O}^u_i=T^3_i+\frac{1}{\sqrt3} T^8_i.
\end{align}
From the weight diagram (Fig.~\ref{Fig: weight diagram}) of
$\bm 8$ representation,
it is clear that this operator has three eigenvalues $1$, 0, and $-1$.
Since the matrix elements of $M_i$ in Eq.\ (\ref{M_i}) are eigenvectors of
$\mathcal{O}^u_i$, we can 
schematically rewrite $M_i$ as
\begin{align}\label{eq: ev of u in M}
M \rightarrow \left(
  \begin{array}{ccc}
    |0\rangle^u & |1\rangle^u & |1\rangle^u \\
    |\mbox{$-1$}\rangle^u &  |0\rangle^u &  |0\rangle^u \\
    |\mbox{$-1$}\rangle^u &  |0\rangle^u &  |0\rangle^u
  \end{array}
\right),
\end{align}
where we have omitted other indices and coefficients
and introduced the eigenvectors as
${\cal O}^u\ket{\mbox{$\pm1$}}^u=\pm \ket{\mbox{$\pm1$}}^u$
and ${\cal O}^u\ket{0}^u=0$.
Performing multiplication of the matrices $M_i$ explicitly,
we see that the eigenstates $\ket{1}^u$ and $\ket{\mbox{$-1$}}^u$
appear in an alternating fashion
in all the product states included in the expansion of $\ket{\Psi}$
if we ignore the states $|0\rangle^u$.
For example, the expansion contains states such as
\[
\cdots\ket{0}^u_{i-2}\ket{1}^u_{i-1}
\ket{\mbox{$-1$}}^u_{i}\ket{0}^u_{i+1}\ket{0}^u_{i+2}
\ket{1}^u_{i+3}\ket{0}^u_{i+4}\ket{\mbox{$-1$}}^u_{i+5}\cdots.
\]%
This structure is exactly the same as the hidden order in
the SU(2) AKLT state.\cite{denNijsRommelse,KennedyTasaki}
We thus define the string order parameter of up quarks as
\begin{equation}
{\cal O}^{\rm str}_{u}=\lim_{|j-k|\rightarrow \infty}\lim_{L\rightarrow \infty}
\langle \Psi |{\cal O}^u_j
\exp\!\left(i\pi \sum_{j\le l<k}{\cal O}^u_l\right)
{\cal O}^u_k | \Psi \rangle
,
\label{eq: string order Ou}
\end{equation}
where the wave function is normalized as $\langle \Psi | \Psi \rangle =1$.
Using the method developed in Refs.~\onlinecite{FannesNachtergaele1,FannesNachtergaele2,KlumperSchadschneiderZittartz}, we explicitly
calculate the string correlation
\begin{align}
\lim_{L\rightarrow \infty}
\langle \Psi |{\cal O}^u_j
\exp\!\left(i\pi \sum_{j\le l<k}{\cal O}^u_l\right)
{\cal O}^u_k | \Psi \rangle
=\frac{1}{4}-\frac{1}{4}\left( -\frac18\right)^{k-j}
\end{align}
and obtain ${\cal O}^{\rm str}_{u}=1/4$.

Similarly, we can define two other flavor operators
\begin{align}
{\cal O}_i^d&=-T^3_i+\frac{1}{\sqrt3} T^8_i,\\
{\cal O}_i^s&=-\frac{2}{\sqrt3} T^8_i.
\end{align}
From the weight diagram it is clear that these two operators
also have the eigenvalues $-1, 0, 1$.
These three flavor operators ${\cal O}^\alpha$ $(\alpha=u,d,s)$
are related to each other
through the $Z_3$ transformation
$U_x$ in Eq.~(\ref{eq: cocycle Z3*Z3})
in the $\bm 3$ representation.
Hence the state $|\Psi\rangle$ has the same string order
${\cal O}^{\rm str}_{u}={\cal O}^{\rm str}_{d}={\cal O}^{\rm str}_{s}=1/4$.

Let us introduce another set of operators defined by
\begin{align}
{\cal O}^a_i&=\frac13 (I^+_i + I^-_i + U^+_i + U^-_i + V^+_i + V^-_i ),\\
{\cal O}^b_i&=\frac13 (\omega^2 I^+_i + \omega I^-_i + \omega^2 U^+_i
+ \omega U^-_i + \omega V^+_i + \omega^2 V^-_i ),\\
{\cal O}^c_i&=({\cal O}^b_i)^\ast
\end{align}
with the raising and lowering operators $I^\pm, U^\pm, V^\pm$
in Eq.~(\ref{eq: raising and lowering ops}).
These three operators are transformed to each other
by the other $Z_3$ transformation $U_y$ in Eq.~(\ref{eq: cocycle Z3*Z3})
in the $\bm 3$ representation.
We can find the eigenvalues of these operators
in the $\bm 8$ representation
by considering the following new basis states of
the $\bm 3$ representation:
\begin{align}
\left(
  \begin{array}{c}
|a\rangle \\
|b\rangle \\
|c\rangle
  \end{array}
\right)=
W\left(
  \begin{array}{c}
|u\rangle\\
|d\rangle\\
|s\rangle
  \end{array}
\right),\ \ \ \
W\equiv\frac{-i}{\sqrt3}
\left(
  \begin{array}{ccc}
    1 & 1 & 1 \\
    1 &  \omega & \omega^2 \\
    1 &  \omega^2 &  \omega
  \end{array}
\right).
\end{align}
The conjugate basis states are given by
\begin{align}
\left(
  \begin{array}{c}
|\bar{a}\rangle \\
|\bar{b}\rangle \\
|\bar{c}\rangle
  \end{array}
\right)=
(W^{-1})^T
\left(
  \begin{array}{c}
|\bar{u}\rangle\\
|\bar{d}\rangle\\
|\bar{s\emph{}}\rangle
  \end{array}
\right).
\end{align}
This follows from a representation of the SU(3) matrix $W$ as
$W=\exp(ix^a \lambda^a)$
and its conjugate representation
$\exp[ix^a (-\lambda^a)^T]=(W^{-1})^T$.
These states ($\ket{a}$, $\ket{b}$, and $\ket{c}$) and their
conjugate states ($\ket{\bar a}$, $\ket{\bar b}$, and $\ket{\bar c}$)
are eigenvectors of the operators ${\cal O}^\alpha_i$ $(\alpha=a,b,c)$.
From Eq.~(\ref{eq: MPS2 SU(3) AKLT a})
we note that the same MPS wave function $\ket{\Psi}$ can be obtained
by replacing the matrix $M$ with $WMW^{-1}$,
which takes the form
\begin{align}
WMW^{-1}&=
W
\begin{pmatrix}
\ket{u} \\
\ket{d} \\
\ket{s} \\
\end{pmatrix}
\begin{pmatrix}
\ket{\bar u} &
\ket{\bar d} &
\ket{\bar s} 
\end{pmatrix}
W^{-1} \n
& ~~~ -\frac{1}{3}\ket{u \bar u+ d \bar d+ s \bar s}1_3,  \n
&=
\begin{pmatrix}
\ket{a} \\
\ket{b} \\
\ket{c} \\
\end{pmatrix}
\begin{pmatrix}
\ket{\bar a} &
\ket{\bar b} &
\ket{\bar c} 
\end{pmatrix}
-\frac{1}{3}\ket{a \bar a+ b \bar b+ c \bar c}1_3.
\end{align}
Being similar to $M$,
the matrix $WMW^{-1}$ can be schematically written as
\begin{align}
WMW^{-1} \rightarrow \left(
  \begin{array}{ccc}
    |0\rangle^a & |1\rangle^a & |1\rangle^a \\
    |\mbox{$-1$}\rangle^a &  |0\rangle^a &  |0\rangle^a \\
    |\mbox{$-1$}\rangle^a &  |0\rangle^a &  |0\rangle^a
  \end{array}
\right), 
\end{align}
where  ${\cal O}^a|\mbox{$\pm 1$}\rangle^a=\pm |\mbox{$\pm 1$}\rangle^a$,
${\cal O}^a|0 \rangle^a=0$,
and we have omitted coefficients and other indices
to simplify presentation.
Since the eigenvectors are arranged in the transformed matrix
in the same way as
in Eq.~(\ref{eq: ev of u in M}),
the eigenvalues of ${\cal O}^a$ also have a hidden order,
which can be measured by the string order parameter
\begin{align}
{\cal O}^{\rm str}_{a}=\lim_{|j-k|\rightarrow \infty}\lim_{L\rightarrow \infty}
\langle \Psi |{\cal O}^a_j \exp\!\left(i\pi \sum_{j\le l<k}{\cal O}^a_l\right)
{\cal O}^a_k | \Psi \rangle.
\end{align}
We obtain
${\cal O}^{\rm str}_{a}={\cal O}^{\rm str}_b={\cal O}^{\rm str}_c=1/4$.

We note that the above string correlations are meaningful indicators
of hidden order in SU(3) symmetric systems, but they are not
necessarily so in $Z_3 \times Z_3$ symmetric ones.
In order for the string correlations to have finite values
in the limit $k-j \to \infinity$,
the largest eigenvalues of the two transfer matrices,
\begin{align}
\mathcal{M}&= \sum_m A^m \tensor (A^m)^*, \\
\widetilde{\mathcal{M}}&=
\sum_{m,n}(e^{i\pi{\cal O}^\alpha})_{mn} A^n \tensor (A^m)^*,
\end{align}
must have the same absolute values.
Otherwise, the string correlations vanish in the limit $k-j\to\infty$.
Since the operator $g=e^{i\pi{\cal O}^\alpha}$ is an element of SU(3),
the eigenvalues of the two transfer matrices coincide 
if the system has the SU(3) symmetry.
In fact, using Eq.~(\ref{eq: symmetry transformation of MPS}),
we have
\begin{align}
\widetilde{\mathcal{M}}&= \sum_{m,n} g_{mn} A^n \tensor (A^m)^* \n
&=\sum_m e^{i\theta_g} U_g^{-1} A^m U_g \tensor (A^m)^* \n
&=e^{i\theta_g} (U_g \tensor 1)^{-1} \mathcal{M} (U_g \tensor 1),
\end{align}
and all the eigenvalues coincide up to a U(1) phase factor.
However, since $e^{i\pi{\cal O}^u_l}$ is not an element of $Z_3 \times Z_3$,
the largest eigenvalues of $\mathcal{M}$ and $\widetilde{\mathcal{M}}$
generally do not coincide for $Z_3 \times Z_3$ symmetric systems.
Therefore the above discussion on the string order
is valid only under the assumption that the system has SU(3) symmetry
(not only the $Z_3 \times Z_3$ symmetry).

Lastly, we examine the relation between boundary states and
hidden symmetry breaking.
While a general theory of a nonlocal $Z_N \times Z_N$ symmetry breaking
is presented in Ref.~\onlinecite{KD-quella-3-13},
we explicitly demonstrate here that a $Z_3 \times Z_3$ symmetry breaking  
takes place in our model by choosing appropriate boundary states.
We consider the MPS under open boundary conditions
\begin{align}
|\Psi(a\bar{u})\rangle=
\frac{1}{3^{L/2} C} v_{a}^\dagger M_1 M_2\cdots M_L v_{\bar u},
\label{eq: MPS symmetry broken state}
\end{align}
where we have chosen the right boundary state $|\bar u\rangle$
at $j=L$ to be $v_{\bar u}=(1,0,0)^T$
and the left boundary state $|a \rangle$ at $j=1$ to be
$v_{a}^\dagger=(1,1,1)/\sqrt3$.
To see a hidden symmetry breaking in $\ket{\Psi(a\bar{u})}$,
we define  the string operators
\begin{align}
O^{\alpha,{\rm str}}_j=
{\cal O}^\alpha_j \exp\left(i\pi \sum_{j<k\le L} {\cal O}^\alpha_k\right)
\label{string operators uds}
\end{align}
for $\alpha=u,d,s$,
where the ``string" operator extends from the right edge ($k=L$)
to the site $j$ in the bulk.
With the right boundary state set to $|\bar u\rangle$,
we obtain the expectation values of the string
operators as
\begin{subequations}
\begin{align}
\langle O^{u,{\rm str}}_j \rangle_{a\bar{u}} &=
-\frac{1}{2} -\frac{1}{4} \left(-\frac{1}{8} \right)^{L-j}
+{\cal O}\biglb(\left(-1/8\right)^{j}\bigrb)
\end{align}
and
\begin{align}
\langle O^{\alpha,{\rm str}}_j \rangle_{a\bar{u}}^{} &=
\frac{1}{2} -\frac{1}{8} \left(-\frac18\right)^{L-j}
+{\cal O}\biglb(\left(-1/8\right)^{j}\bigrb)
\end{align}
\end{subequations}
for $\alpha=d,s$.
Here
$\langle O\rangle_{a\bar{u}}:=
\langle\Psi(a\bar{u})|O|\Psi(a\bar{u})\rangle$.
We note that these results hold for arbitrary left boundary states at $j=1$.
This indicates that a hidden $Z_3$ symmetry is broken in the bulk
in the direction selected by the applied boundary field (state).

Similarly, we consider the other set of $Z_3$ string operators
defined by
\begin{align}
O_j^{\alpha,{\rm str}}=
\exp\!\left({i\pi \sum_{1 \le k<j} {\cal O}^\alpha_k}\right){\cal O}^\alpha_j
\label{string operators abc}
\end{align}
for $\alpha=a,b,c$,
where the ``string" operators connect the left edge ($k=1$)
to the site $j$ in the bulk.
Since the left boundary state is set to $|a \rangle$, we obtain
\begin{subequations}
\begin{equation}
\langle O_j^{a,{\rm str}} \rangle_{a\bar{u}} =
\frac{1}{2} +\frac{1}{4} \left(-\frac18\right)^{j-1}
+{\cal O}\biglb(\left(-1/8\right)^{L-j}\bigrb),
\end{equation}
and
\begin{equation}
\langle O_j^{\alpha,{\rm str}}\rangle_{a\bar{u}} =
-\frac{1}{2} +\frac{1}{8} \left(-\frac18\right)^{j-1}
+{\cal O}\biglb(\left(-1/8\right)^{L-j}\bigrb)
\end{equation}
\end{subequations}
for $\alpha=b,c$.
These results also hold for arbitrary right boundary states.
This indicates that another hidden $Z_3$ symmetry
is broken in the bulk by selecting the left boundary state.
We note that the string operators defined
in Eqs.\ (\ref{string operators uds})
and (\ref{string operators abc}) take different expectation values
for the two ground-state wave functions $\ket{\Psi}$
and $\ket{\widetilde{\Psi}}$ of $H_3$,
as they are operators without inversion symmetry.
On the other hand, the string correlation function in
Eq.\ (\ref{eq: string order Ou}) takes the same value for the two states.

We can perform the $Z_3$ rotation of the ``symmetry broken" state
[Eq.~(\ref{eq: MPS symmetry broken state})]
by applying the $Z_3$ transformations $x$ and $y$ given in 
Eq.\ (\ref{x and y}),
under which the
boundary states are transformed projectively with $U_x$ and $U_y$
as in Eq.~(\ref{eq: transformation of boundary states}).
It turns out that
the two $Z_3$ transformations $x$ and $y$ act differently on the
two sets of the expectation values of the string operators
$\langle O_j^{\alpha,{\rm str}}\rangle_{a\bar{u}}$.
In fact, $U_x$ causes the $Z_3$ rotation only among
the right boundary states $\{|\bar u\rangle, |\bar d\rangle, |\bar s\rangle\}$
and does not change the left boundary states $|\alpha \rangle$ ($\alpha=a,b,c$)
up to phase factors
(e.g., $v_{a}^\dagger U_x^{-1}=v_{a}^\dagger$).
Thus an action of $x$ interchanges the values of
$\langle O_j^{\alpha,{\rm str}}\rangle$ for $\alpha=u,d,s$,
without changing those of $\alpha=a,b,c$.
Similarly, $U_y$ causes the $Z_3$ rotation among
the left boundary states $\{|a\rangle, |b\rangle, |c\rangle\}$
without changing the right boundary states $|\bar \alpha\rangle$
($\alpha=u,d,s$) up to phase factors
(e.g., $U_y v_{\bar u}=v_{\bar u}$)
so that $y$ interchanges the values of
$\langle O_j^{\alpha,{\rm str}}\rangle$ for $\alpha=a,b,c$ only.
Therefore, two independent $Z_3$ symmetries
(i.e., hidden $Z_3\times Z_3$ symmetry) are broken
in the SU(3) AKLT state with boundary vectors given in
Eq.\ (\ref{eq: MPS symmetry broken state}).

\section{DMRG results\label{sec: DMRG}}

In Sec.\ \ref{sec: SU(3) AKLT model} we have constructed
the SU(3) AKLT Hamiltonian, which turns out to be
a special case of the SU(3) bilinear-biquadratic Hamiltonian.
In this section we study the latter Hamiltonian
by means of the iDMRG\cite{white-92,white-93,McCulloch-08}
and obtain its ground-state phase diagram.

\subsection{SU(3) bilinear-biquadratic model}
We study the SU(3) bilinear-biquadratic Hamiltonian
\begin{equation}
H_{\theta} = \sum_i [\cos \theta \, \bm{T}_i \cdot \bm{T}_{i + 1} 
+ \sin \theta \, ( \bm{T}_i \cdot \bm{T}_{i + 1} )^2],
\label{eq: SU(3) Hamiltonian}
\end{equation}
where $\bm{T}_i$ are su(3) generators in $\bm8$ representation.
The SU(3) AKLT Hamiltonian
(\ref{eq: SU(3) AKLT Hamiltonian}) 
is a special case of $H_\theta$ at $\theta = {\rm arctan}(2/9)$
(up to an overall numerical factor).
Obviously, $H_\theta$
is invariant under any SU(3) transformation, thereby
invariant under the symmetry group $Z_3 \times Z_3$.

\begin{figure}[tb]
\includegraphics[width=\linewidth]{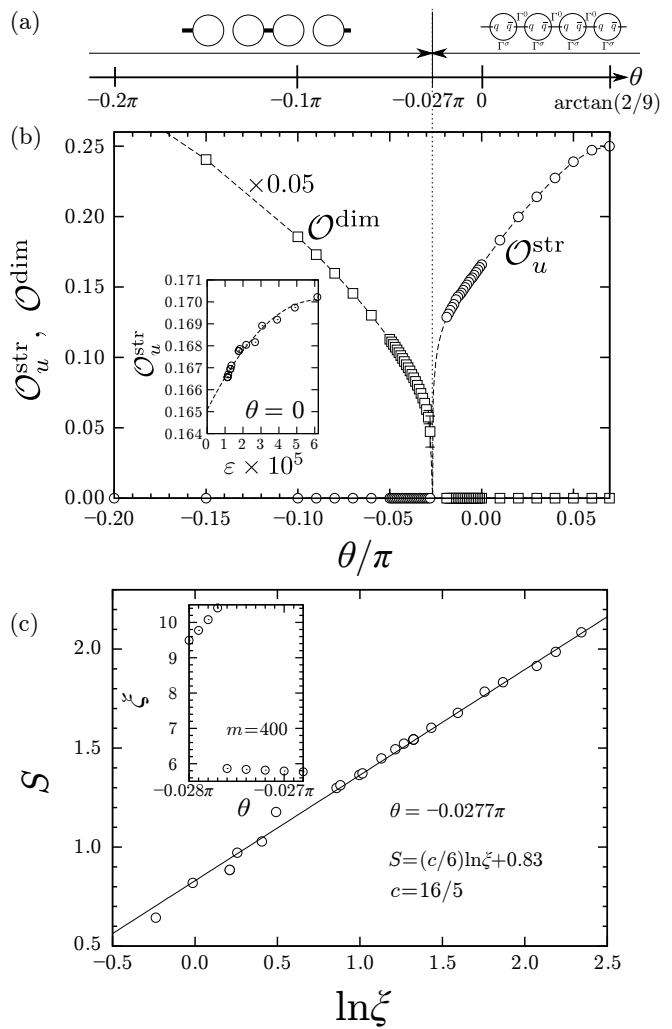}
  \caption{
(a) Phase diagram of the SU(3) bilinear-biquadratic Hamiltonian.
(b) The string order $\mathcal{O}^{\rm str}_{u}$
[Eq.~(\ref{eq: string order Ou})] 
and the dimer order $\mathcal{O}^{\rm dim}$ [Eq.~(\ref{eq:dimer})]
as functions of $\theta$. 
The number of kept states $m$ in the iDMRG calculation is
taken up to $m = 200$.
Plotted are the values of data extrapolated
to the limit of vanishing truncation errors
$\varepsilon\to0$.
The inset shows an example of the extrapolation
for $\mathcal{O}^{\rm str}_{u}$
at $\theta = 0$ with a quadratic fit.
The dotted line indicates the phase boundary
between the $\mathbb{Z}_3$ SPT phase and the dimer phase
at $\theta_c \approx -0.027\pi$. 
The broken curves are to guide the eye. 
(c) The entanglement entropy $S$ [Eq.~(\ref{eq:ee})] 
as a function of the correlation length $\xi$ [Eq.~(\ref{eq:xi})] 
at $\theta = -0.0277\pi$,
where $\xi$ takes a maximum value for $m = 400$ as shown in the inset. 
The solid line shows the asymptotic behavior of
the entanglement entropy 
at criticality
with the central charge $c=16/5$. 
}
\label{fig:pd}
\end{figure}
Figure~\ref{fig:pd}(a) shows the phase diagram of 
the SU(3) bilinear-biquadratic Hamiltonian [Eq.~(\ref{eq: SU(3) Hamiltonian})]
with $\theta$ in the parameter range $[-0.2\pi, \arctan(2/9)]$. 
We have the $\mathbb{Z}_3$ SPT phase for $\theta\gtrsim -0.027\pi$
and a dimer phase for $\theta \lesssim -0.027\pi$.
We note that for $\theta\gtrsim -0.027\pi$,
one of the two SPT states shown in Fig.~\ref{Fig: SU(3) AKLT wave function}
is spontaneously selected in the iDMRG calculation,
because a macroscopic superposition of two SPT states
needs twice as many kept states in an MPS representation 
as either one of two SPT states of the same energy does.
In the dimer phase, the translation symmetry is spontaneously broken
and SU(3) singlet dimers are formed by $\bm{8}$ states
from neighboring sites,
as schematically shown in Fig.~\ref{fig:pd}(a).
These two phases are distinguished by the string order
$\mathcal{O}^{\rm str}_{u}$ in Eq.~(\ref{eq: string order Ou}) 
and a dimer order $\mathcal{O}^{\rm dim}$ defined by
\begin{equation}
\mathcal{O}^{\rm dim} =
\left|
 \bra{\Psi} \bm{T}_1\cdot\bm{T}_{2} - \bm{T}_2\cdot\bm{T}_{3} \ket{\Psi}
\right|
\label{eq:dimer}
\end{equation}
for an infinite MPS with two-site periodicity.
We find a finite string order ($\mathcal{O}^{\rm str}_{u} > 0$)
and no dimer order ($\mathcal{O}^{\rm dim} = 0$)
in the $\mathbb{Z}_3$ SPT phase,
while $\mathcal{O}^{\rm str}_{u} = 0$ and $\mathcal{O}^{\rm dim} > 0$
in the dimer phase, as shown in Fig.~\ref{fig:pd}(b).

We determine the phase transition point $\theta_{c}$ as follows.
We first perform extrapolations of $O^{\rm str}_{u}, O^{\rm dim}$
with the truncation error~\cite{white-92,white-93} for each value of $\theta$. 
The truncation error 
is defined by $\varepsilon = \sum_{i=m+1}^{8m} w^{2}_{i}$
in the final numerical iteration of iDMRG,
where $m$ is the number of kept internal states. 
Numerically accurate estimates for the order parameters can be obtained
by taking extrapolations to $\varepsilon \rightarrow 0$.
Following Ref.~\onlinecite{white07PRL},
we fit order parameters with a quadratic form of $\varepsilon$
as shown in the inset of Fig.~\ref{fig:pd}(b).
Errors of data points in Fig.~\ref{fig:pd}(b) are smaller than
the symbols except for those at $\theta=-0.027\pi$ and $-0.028\pi$.
However, we did not obtain reasonable quadratic fits for the data of
$m \leq 200$ in the range $-0.027\pi \leq \theta \leq -0.020\pi$.
We then fit extrapolated values of
$O^{\rm str}_{u}$ and $O^{\rm dim}$
with power-law functions
$f_{\rm str} = a(\theta - \theta_c)^{\beta}$ and
$f_{\rm dim} = a'(\theta'_c - \theta)^{\beta'}$
in the vicinity of the phase boundary, where
$a$, $a'$, $\theta_{c}$, $\theta'_{c}$, $\beta$ and $\beta'$
are fitting parameters.
We obtain $\theta_{c}$ and $\theta'_{c}$ 
from fitting of the data
in the range $-0.019\pi \leq \theta \leq -0.01\pi$
and $-0.037\pi \leq \theta \leq -0.028\pi$ as
\begin{align}
\theta_c/\pi&=-0.028^{+0.002}_{-0.003}, &
\theta'_c/\pi&=-0.027^{+0.005}_{-0.001}.
\label{eq: theta_c}
\end{align}%
We find that the fitting parameters vary when
we change fitting ranges even in the region
where we obtain good extrapolations of $O^{\rm str}_{u}$
and $O^{\rm dim}$ to $\varepsilon \to 0$,
because these regions are not sufficiently close to
the critical point.
Thus 
the parameters obtained from our numerical calculation may not be very
reliable by themselves.
However, if we assume that the phase transition takes place at a single
point,
i.e., $\theta_c = \theta'_c$, then
the combination of the estimates for $\theta_c$ and $\theta'_c$ 
can provide a more reliable estimate for the critical
point $\theta_c$.
Furthermore, the critical point should be located in between
the regions where either of the two order parameters is finite.
We find the overlapping region to be $\theta_c/\pi = -0.027\pm0.001$ 
from $\theta_c$ and $\theta_c'$ in Eq.\ (\ref{eq: theta_c}).
Thus, we conclude that the critical point
is at $\theta_c/\pi = -0.027\pm0.001$.

\subsection{Criticality at the phase transition}
We study criticality at the phase transition between
the $\mathbb{Z}_3$ SPT phase and the dimer phase,
using the scaling of the entanglement entropy of a bipartition.
The entanglement entropy is given by
\begin{align}
S = {\rm Tr} [-\rho_{\rm L} \ln \rho_{\rm L} ],
\end{align}
where $\rho_{\rm L}$ is a reduced density matrix given by an integral
of the density matrix over the Hilbert space of the right chain as
$\rho_{\rm L} = {\rm Tr}_{R}\ket{\Psi} \bra{\Psi}$. 
Critical points of 1D quantum systems
are described by conformal field theories.
In the vicinity of a critical point,
the entanglement entropy $S$  
increases logarithmically with the correlation length $\xi$ of the system as 
\begin{equation}
S \sim \frac{c}{6} \ln \xi + S_0 ,
\label{eq:ee}
\end{equation}
where $c$ is the central charge of the underlying conformal field theory
and $S_0$ is a nonuniversal constant.\cite{vidal-entanglement03,calabrese04}
For a wave function of the MPS form,
the correlation length $\xi$ is given by~\cite{Wolf06} 
\begin{equation}
\xi = \frac{1}{\ln |\mu_1/\mu_2|},
\label{eq:xi}
\end{equation}
where $\mu_1$ and $\mu_2$ are dominant and subdominant eigenvalues
of the two-site transfer matrix 
\begin{equation}
\mathcal{M}_2 = \sum_{m_1,m_2} A_1^{m_1} A_2^{m_2} \otimes (A_1^{m_1} A_2^{m_2})^{*} .
\label{eq:m2}
\end{equation}
Here we consider the two-site transfer matrix rather than
the single-site transfer matrix 
in order to obtain a unique dominant eigenvalue $\mu_1$ in the dimer phase
where wave functions break the translation symmetry
and have a period of two sites.\cite{ueda11}
In Fig.~\ref{fig:pd}(c), we show the entanglement entropy plotted
as a function of $\xi$ at $\theta=-0.0277\pi$
which is the peak position of $\xi$ 
for $m=400$ [see the inset of Fig.~\ref{fig:pd}(c)]
This peak position is consistent with the estimate
$\theta_c/\pi = -0.027\pm0.001$.
From this analysis,
we find that the entanglement entropy $S$ fits well
to the formula of Eq.\ (\ref{eq:ee})
with $c=16/5$.

Since the bilinear-biquadratic Hamiltonian $H_\theta$ has
SU(3) symmetry, we expect that the critical point should be
described by an SU(3) Wess-Zumino-Witten (WZW) model with some
level $k$, i.e., SU(3)$_k$ WZW model.
The central charge of the SU$(N)_{k}$ WZW model is
\begin{align}
c=\frac{k(N^2-1)}{k+N}.
\end{align}
It is well known that
the transition between the Haldane phase and the dimer phase in the
SU(2) bilinear-biquadratic model for $S=1$ spins 
is described by the SU(2)$_2$ WZW model,
which has $c=3/2$.\cite{Affleck86,Affleck-Haldane87,alcaraz-martins88}
It is also known that the SU(3) bilinear-biquadratic model
in $\bm{6}$ representation
shows a criticality described by the SU(3)$_2$ WZW
model ($c=16/5$).\cite{andrei-johannesson84,fuhringer08}
Thus the central charge $c=16/5$ observed in Fig.~\ref{fig:pd}(c)
suggests that the criticality
between the $\mathbb{Z}_3$ SPT phase and the dimer phase
is also described by the SU(3)$_2$ WZW model.

Assuming the SU(3)$_2$ WZW criticality,
we might speculate
the critical exponent for the dimer order $\beta'$
as follows.
The effective action around the critical point is
\begin{align}
S(g)=S_{\t{WZW}}(g)+ t\int d^2 x ~ \Phi ,
\label{eq: S(g)}
\end{align}
where $S_{\t{WZW}}(g)$ is the action of the SU(3)$_2$ WZW model,
$g$ is an SU(3) matrix field, and
the coupling constant $t\propto\theta-\theta_c$.
The operator $\Phi$ is a relevant operator 
that is an SU(3) symmetric scalar and 
respects translation symmetry.
The WZW model for our system of SU(3) spins in the adjoint representation
is presumably obtained in the strong-interaction limit of
a Hubbard model of fermions with three flavors and two colors
(labeling a quark and an antiquark).
The quark fermions are 1/3-filled and the antiquark fermions are
2/3-filled.
We speculate that a relevant operator permitted by the symmetry is
unique and is the primary field $\Phi$ corresponding to the adjoint
representation, whose scaling dimension
is\cite{Affleck86,Affleck-Haldane87,Affleck-SU(n)88,DiFrancesco}
\begin{align}
x=\frac{2C(\bm{N^2-1})}{N+k}=\frac{6}{5}.
\end{align}
If so, then from the scaling equation
\begin{align}
\frac{dt}{d \ln L}=(2-x) t,
\end{align}
we find that the correlation length $\xi$ diverges as
\begin{align}
\xi &\propto |\theta-\theta_c|^{-\nu}, &
\nu&=\frac{1}{2-x} =\frac{5}{4}. 
\end{align}
By an analogy with the SU(2) case,%
\cite{Affleck86,Affleck-Haldane87}
we speculate that 
the dimer order is given by the operator $\t{tr}\,g$
in the WZW model,
whose scaling dimension is 
\begin{align}
x_{\t{dim}}= \frac{2C(\bm{N})}{N+k}=\frac{8}{15}.
\end{align} 
Since the dimer order parameter scales with the correlation length as
$\mathcal{O}^{\rm dim} \propto 1/\xi^{x_\t{dim}}$,
the critical behavior of the dimer order $\beta'$ 
is presumably given by
\begin{align}
\mathcal{O}^{\rm dim} &\propto (\theta_c-\theta)^{\beta'}, &
\beta'&=\nu x_\t{dim} =\frac{2}{3}.
\label{eq: beta'}
\end{align}
Unfortunately, we could not obtain reliable
estimates for the critical exponents $\nu$ and $\beta'$
from our numerical data presented in Fig.~\ref{fig:pd}(b),
because, with the limited number of kept states and CPU time,
our iDMRG calculation did not reach
sufficiently good convergence
for the dimer and string order 
parameters in the very vicinity of the critical point.

\subsection{Entanglement spectrum}
\begin{figure}[tb]
\includegraphics[width=\linewidth]{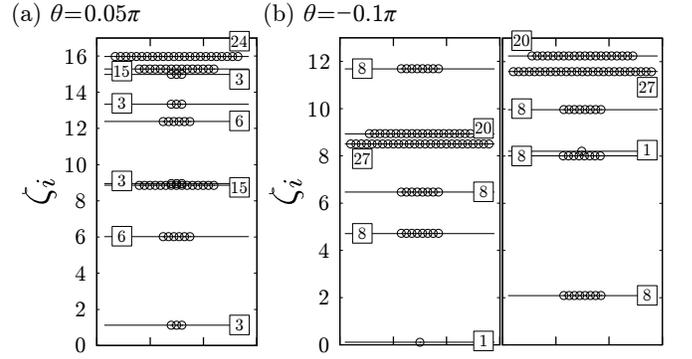}
  \caption{
Entanglement spectrum of the SU(3) bilinear-biquadratic model
[Eq.~(\ref{eq: SU(3) Hamiltonian})] for
(a) $\theta/\pi = 0.05$ in the $\mathbb{Z}_3$ SPT phase
and (b) $\theta/\pi = -0.1$ in the dimer phase.
The left/right panel in (b) shows entanglement spectrum
of the chain divided without/with cutting a singlet dimer.
The numbers enclosed in squares indicate the degeneracy of
multiplets in the entanglement spectra.
}
\label{fig:es}
\end{figure}
We study the entanglement spectrum in the $\mathbb{Z}_3$ SPT phase 
and the dimer phase.
The entanglement spectrum $\{\zeta_i\}$ is defined,
via the entanglement Hamiltonian\cite{li-haldane-08} 
\begin{subequations}
\begin{equation}
H_{E} = - \ln \rho_L
 = \sum_{i} \zeta_i \ket{\psi^{L}_{n}}_i ~\bra{\psi^{L}_n}_i,
\end{equation}
by
\begin{equation}
\zeta_i = - \ln w^2_i
\end{equation}
\end{subequations}
where left singular vectors $\ket{\psi^{L}_{n}}_i$ and
singular values $w_i$ are introduced in Eq.~(\ref{SVD}).
The entanglement spectrum in the $\mathbb{Z}_3$ SPT phase shown
in Fig.~\ref{fig:es}~(a) has the degeneracy in multiples of three. 
This signals that the ground state is in the SPT phase
protected by $Z_3 \times Z_3$ symmetry. 
In the dimer phase, the entanglement spectrum depends on the position 
where we cut the spin chain,
because the ground state breaks the translation symmetry
[Fig.~\ref{fig:es}~(b)]. 

\begin{figure}[tb]
\begin{center}
\includegraphics[width=0.5\linewidth]{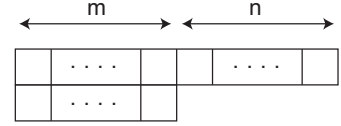}
\end{center}
\caption{
Young tableau of the $[n,m]$ representation of SU(3).
}
\label{Fig: Young tableau}
\end{figure}
Since $H_\theta$ has the SU(3) symmetry,
eigenstates of the entanglement Hamiltonian form SU(3) multiplets.
Irreducible representations of SU(3) are specified by
two integers as $[n,m]$,
with the Young tableau shown in Fig.~\ref{Fig: Young tableau}.
The dimension of the $[n,m]$ representation of SU(3)
is\cite{geourgi-Lie-algebras}
\begin{align}
D(n,m)&=\frac{1}{2}(n+1)(m+1)(n+m+2),
\label{eq:dnm}
\end{align}
and for the first few irreducible representations
$
D(0,0)=1,~ D(1,0)=3,~ 
D(1,1)=8,~ D(2,0)=6,~
D(2,1)=15,~ D(2,2)=27,~ 
D(3,0)=10,~ D(3,1)=24.
$
These dimensions agree with the degeneracies found in
the entanglement spectrum in Fig.~\ref{fig:es}, 
except for the 20-fold degeneracy in the dimer phase,
which might be attributed to an extra doubling due to a dimer formation.

\section{Building SU(3) and SU(4) models with SU(2) spin chains
\label{sec: SU(3) and SU(4)}}
The Haldane phase of the antiferromagnetic spin-1 chain is
closely related to the dimer phase of spin-$\frac12$ chains.\cite{Hida92}
In fact, the dimerized ground state of the spin-$\frac12$ Heisenberg chain
with alternating exchange coupling ($J'<J$ and $J>0$)
becomes the AKLT state of the spin-1 chain in the limit $J'\to-\infty$.
In this section we relate
AKLT states to dimerized states for the SU(3) and SU(4)
versions of the AKLT model.

\subsection{Spin quadrupole operators and the SU(3) AKLT model\label{sec: spin quadrupolar}}

We consider SU(2) spin $S=1$ chains with staggered biquadratic couplings
and show that their ground states are in the dimer phase
which is adiabatically
connected to
a $\mathbb{Z}_3$ SPT phase of the SU(3) AKLT model
introduced in Sec.\ \ref{sec: SU(3) AKLT model B}.

\subsubsection{$S=1$ spin chains with staggered biquadratic couplings}

For the three-dimensional Hilbert space of $S=1$,
we find it convenient to take the basis\cite{Lauchli06,penc-lauchli11}
\begin{align}
\ket{x}&=i\frac{\ket{1}-\ket{{-1}}}{\sqrt2}, &
\ket{y}&=\frac{\ket{1}+\ket{{-1}}}{\sqrt2}, &
\ket{z}&=-i\ket{0}, &
\label{eq: S=1 basis for su(3) ops}
\end{align}
rather than the basis $\ket{n}$ diagonalizing $S^z$,
$S^z\ket{n}=n\ket{n}$ with $n=-1,0,1$.
In the new basis
the spin operators in Eq.\ (\ref{S=1 spin}) are written as
\begin{align}
(S^\alpha)_{\beta\gamma}
=-i\epsilon_{\alpha\beta\gamma}\ket{\beta}\langle\gamma|
\qquad(\alpha,\beta,\gamma=x,y,z),
\end{align}
where $\epsilon_{\alpha\beta\gamma}$ is a totally antisymmetric tensor
with $\epsilon_{xyz}=+1$.
We define spin quadrupole operators:
\begin{align}
\Vec{Q}=
\begin{pmatrix}
Q^{x^2-y^2} \\
Q^{3 z^2-r^2} \\
Q^{yz} \\
Q^{zx} \\
Q^{xy} \\
\end{pmatrix}
:=
\begin{pmatrix}
(S^x)^2-(S^y)^2 \\
\frac{1}{\sqrt{3}} [3(S^z)^2-2] \\
S^y S^z + S^z S^y \\
S^z S^x + S^x S^z \\
S^x S^y + S^y S^x \\
\end{pmatrix}
.
\end{align}
In the basis of Eq.~(\ref{eq: S=1 basis for su(3) ops}),
the spin and quadrupole operators
are written as
\begin{subequations}
\label{eq: S=1 ops and SU(3) ops}
\begin{align}
\Vec{S}&=(\hat\lambda_7,-\hat\lambda_5,\hat\lambda_2)^T, \\
\Vec{Q} &=
(-\hat\lambda_3,\hat\lambda_8,-\hat\lambda_6,-\hat\lambda_4,-\hat\lambda_1)^T,\\
\hat \lambda_a &=\sum_{\alpha,\beta=x,y,z}
(\lambda_a)_{\alpha\beta}\ket{\alpha}\langle\beta|, 
\end{align}
\end{subequations}%
where $\lambda_a$ ($a=1,\dots,8$) are the Gell-Mann matrices
in Eq.\ (\ref{Gell-Mann matrices}).
Thus the spin and quadrupole operators together give a set of su(3) generators
($T^a=\hat \lambda_a/2$) in the fundamental representation \textbf{3} 
if we multiply them with factors $\pm 1/2$.
The quadratic Casimir operator from the operators of the $i$th and $j$th
sites is then given by
\begin{align}
\sum^8_{a=1} \hat \lambda_a(i) \hat \lambda_a(j) &=
\Vec{S}_i \cdot \Vec{S}_j + \Vec{Q}_i \cdot \Vec{Q}_j \n
&= 2 \Vec{S}_i \cdot \Vec{S}_j +2 (\Vec{S}_i \cdot \Vec{S}_j)^2 - \frac{8}{3}.
\end{align}
Here we have used the identity
\begin{align}
\Vec{Q}_i \cdot \Vec{Q}_j
= 2(\Vec{S}_i \cdot \Vec{S}_j)^2 +  \Vec{S}_i \cdot \Vec{S}_j
 - \frac{2}{3}S^2(S+1)^2.
\end{align}
The su(3) generators for the conjugate representation $\bm\bar{\bm 3}$,
$\hat{\bar{\lambda}}_a/2$, are obtained from
the fundamental representation \textbf{3} as
\begin{align}
\hat{\bar{\lambda}}_a = - \hat{\lambda}_a^*
= \sum_{\alpha,\beta} -(\lambda_a^*)_{\alpha\beta}\ket{\alpha}\langle\beta|.
\end{align}
The su(3) generators in the $\bm\bar{\bm 3}$ representation
are related to $S=1$ spin operators
as in Eqs.\ (\ref{eq: S=1 ops and SU(3) ops}),
where we replace $\hat\lambda_a$ with $\hat{\bar{\lambda}}_a$
and replace spin dipole and quadrupole operators as
\begin{align}
\bm{S} &\to \bm{S}, &
\bm{Q} &\to -\bm{Q}.
\end{align}
Therefore, the quadratic Casimir operator constructed from
$\bm{3}$ at the $i$th site and $\bm\bar{\bm 3}$ at the $j$th site reads
\begin{align}
\sum_a \hat \lambda_a(i) \hat{\bar{\lambda}}_a(j) &=
\Vec{S}_i \cdot \Vec{S}_j - \Vec{Q}_i \cdot \Vec{Q}_j \n
&= -2 (\Vec{S}_i \cdot \Vec{S}_j)^2 + \frac{8}{3}.
\label{Casimir = biquadratic}
\end{align}

Let us consider the following $S=1$ spin chain
with alternating biquadratic interactions:
\begin{equation}
\mathcal{H}_3=-\sum_{i}\left[ J' (\bm{S}_{i,1} \cdot \bm{S}_{i,2})^2
+J (\bm{S}_{i,2} \cdot \bm{S}_{i+1,1})^2 \right] .
\label{eq: S=1 Hamiltonian with staggered biquadratic terms}
\end{equation}
Each unit cell has two $S=1$ spins ($\bm{S}_{i,1}$ and $\bm{S}_{i,2}$),
and we can regard one of them ($\bm{S}_{i,1}$) as
in the $\bm 3$ representation and the other ($\bm{S}_{i,2}$)
as $\bm\bar{\bm 3}$.
From Eq.\ (\ref{Casimir = biquadratic}) we see that the
Hamiltonian is a sum of the quadratic Casimir operators
of $\bm 3$ and $\bm\bar{\bm 3}$ representations from neighboring sites.
The product of $\bm 3$ and $\bm\bar{\bm 3}$ representations
is split by the biquadratic coupling into
an octet and a singlet,
$\bm 3 \tensor \bm\bar{\bm 3} = \bm 8 \oplus \bm 1$.
A negative sign of $J'$ favors an octet, and a positive $J$ favors a singlet.
Therefore we expect that the ground state of $\mathcal{H}_3$ with
$J'<0$ and $J>0$ should be adiabatically connected to the MPS wave function 
(\ref{eq: MPS SU(3) AKLT}) of the SU(3)
AKLT model introduced in Sec.\ \ref{sec: SU(3) AKLT model B}.
To verify this conjecture,
we numerically study the ground-state properties of
the Hamiltonian $\mathcal{H}_3$ using the iDMRG method
and determine the phase diagram as a function of $J'/J$
below.

\subsubsection{DMRG results for the Hamiltonian $\mathcal{H}_3$}

\begin{figure}[tb]
\includegraphics[width=\linewidth]{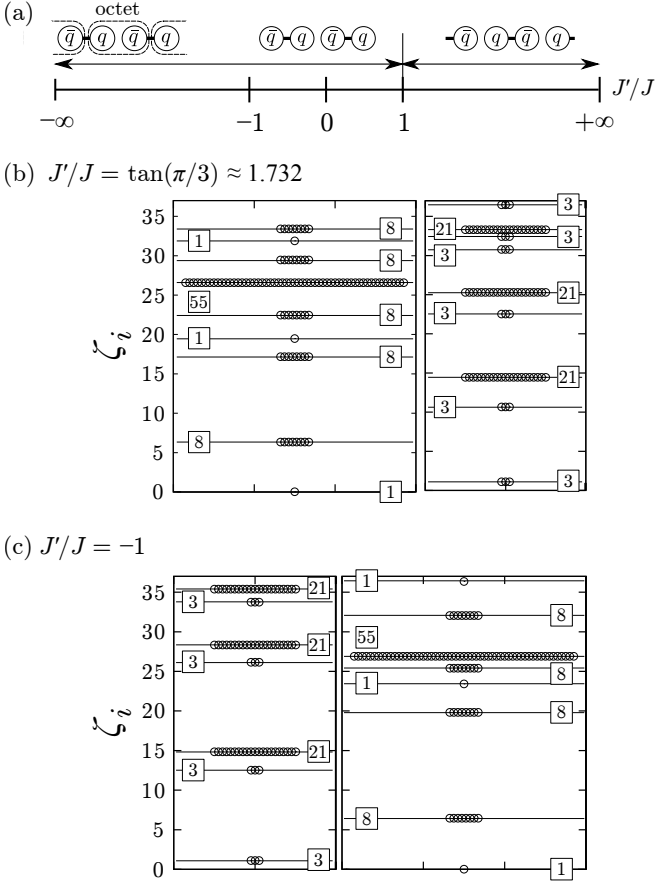}
  \caption{
(a) Phase diagram of the Hamiltonian $\mathcal{H}_3$
as a function of $J'/J$ with $J>0$.
Entanglement spectrum of the ground-state wave function for
(b) $J'/J=\tan(\pi/3)$ and (c) $J'/J=-1$. 
In (b) and (c) the left (right) panels 
show the spectrum 
when we cut the spin chain at a bond between unit cells 
(within a unit cell),
i.e., at a bond of $\bar{q}$--$q$ ($q$--$\bar{q}$),
where $q$ and $\bar q$ are $\bm 3$ and $\bm{\bar{3}}$ states. 
The numbers enclosed in squares indicate
the degeneracy of multiplets in the entanglement spectrum. 
}
\label{fig:bq_alt}
\end{figure}

We obtain the ground state of the Hamiltonian $\mathcal{H}_3$
using the iDMRG method.
We show its phase diagram
as a function of $J'/J$ with $J>0$ in Fig.~\ref{fig:bq_alt}(a).
The phase diagram has two types of dimer phases
which are separated at the point of uniform coupling $J'/J=1$.
At this point the ground state is spontaneously dimerized
and twofold degenerate,
and the energy spectrum is gapped.\cite{barber89,klumper89,Xian93}
Away from the transition point $J'/J=1$,
one of the two dimerization patterns is selected for
the ground state.

The patterns of singlet dimers in the two dimer phases are different
as shown in Fig.~\ref{fig:bq_alt}(a)
and reflected in the entanglement spectra
in Fig.~\ref{fig:bq_alt}(b) and (c).
When the spin chain is divided at a dimerized (singlet) bond,
the degeneracy of the lowest multiplet in the entanglement spectrum
is three as shown in the right panel of Fig.~\ref{fig:bq_alt}(b)
and the left panel of Fig.~\ref{fig:bq_alt}(c).
On the other hand, 
when the spin chain is divided at a un-dimerized bond
the degeneracy of the lowest multiplet in the entanglement spectrum
is one as shown in the left panel of Fig.~\ref{fig:bq_alt}(b) 
and the right panel of Fig.~\ref{fig:bq_alt}(c).
We note that the ground state at $J'/J = J_0$
with $J_0>0$ and $J_0\ne1$
can be transformed to the ground state at $J'/J = 1/J_0$ 
by site-centered inversion. 
Since the site-centered inversion swaps two types of dimerized bonds
$q$--$\bar{q}$ and $\bar{q}$--$q$,
two patterns of singlet dimers are interchanged 
and so are the two entanglement spectra obtained from two ways of
cutting the spin chain.
We also confirm that the degeneracy of the lowest multiplet in the
entanglement spectrum for the dimer phase of $J'/J \leq 1$
remains the same as shown in Fig.~\ref{fig:bq_alt}(c) and
does not change, in particular, across the point $J'/J = 0$,
at which the spin chain is decomposed into a collection of 
SU(3) singlets of $\bar{q}$--$q$.
We note that 
the 21-fold degeneracy and the 55-fold degeneracy in Fig.~\ref{fig:bq_alt}(b)
and (c) correspond to $D(5,0), D(0,5)$ and
$D(9,0), D(0,9)$ of Eq.~(\ref{eq:dnm}), respectively.

When the spin chain is divided at a bond between unit cells,
the entanglement spectrum for $J'/J<1$
shows the degeneracy in multiples of three,
as in the entanglement spectrum in
the $\mathbb{Z}_3$ SPT phase in Fig.~\ref{fig:es}(a).
This indicates that the ground-state wave function of
the Hamiltonian $\mathcal{H}_3$ with $J'/J<1$ 
is adiabatically connected to the SU(3) AKLT state $\ket{\Psi}$ of
the $\mathbb{Z}_3$ SPT phase ($1\in\mathbb{Z}_3$)
discussed in Sec.~\ref{sec: SU(3) AKLT model}.
In the limit of $J'/J\to -\infinity$,
$q$ and $\bar{q}$ states in each unit cell are projected onto
an octet,
and a pair of $\bar{q}$ and $q$ from neighboring unit cells
form a singlet state,
which clearly indicates the connection to the SU(3) AKLT wave function
in Eq.~(\ref{eq: MPS SU(3) AKLT}).

\subsubsection{A model of $S=1$ spins reducing to the SU(3) bilinear Hamiltonian in the strong-coupling limit}

We shall introduce a slightly different 1D Hamiltonian of $S=1$ spins
which should belong to the same $\mathbb{Z}_3$ SPT phase
and reduces, in the strong-coupling limit, to the SU(3) bilinear-biquadratic
Hamiltonian $H_\theta$
at $\theta=0$.
To motivate, we begin with the Hamiltonian $\mathcal{H}_3$
in the strong-coupling limit $-J'/J\gg1$, where the low-energy multiplets
in each unit cell are an octet, the $\bm 8$ representation.
The effective Hamiltonian is then obtained by
writing the remaining biquadratic interactions,
$-J(\bm{S}_{i,2}\cdot\bm{S}_{i+1,1})^2$,
in the subspace of the $\bm 8$ representations.
In each unit cell the octet can be written as
\begin{align}
\ket{\lambda_a}=\frac{1}{\sqrt2}
\sum_{\alpha,\beta}(\lambda_a)_{\alpha\beta}
\ket{\alpha}\tensor \ket{\beta},
\end{align}
where $\ket{\alpha}$ and $\ket{\beta}$ are $S=1$ spin states
in the $\bm 3$ and $\bm\bar{\bm 3}$ representations.
Thus the matrix elements of $\hat{\lambda}_a$ and
$\hat{\bar{\lambda}}_a$ for the octet states
$\ket{\lambda_b}$
are given by
\begin{align}
\label{matrix elements}
\begin{aligned}
\langle \lambda_c | \hat \lambda_a \ket{\lambda_b}
&=\frac{1}{2}\t{tr}(\lambda_a\lambda_b\lambda_c) = d_{abc}+if_{abc}, \\
\langle \lambda_c | \hat{\bar{\lambda}}_a \ket{\lambda_b}
&=-\frac{1}{2}\t{tr}(\lambda_c\lambda_b\lambda_a) = -d_{abc}+if_{abc}.
\end{aligned}
\end{align}
Here we have used the formula
\begin{align}
\lambda_a\lambda_b &=\frac{2}{3}\delta_{ab} 1_3 +
(d_{abc}+if_{abc})\lambda_c
\end{align}
with a symmetric tensor $d_{abc}$ and the structure constant
$f_{abc}$ which is an antisymmetric tensor.
Noting that su(3) generators in the $\bm 8$ representation
are given by $T^i_{jk}=-if_{ijk}$, we see that the matrix elements in
Eqs.\ (\ref{matrix elements}) have additional
contributions of the symmetric tensor $d_{abc}$ which cannot be
written in terms of su(3) generators.
We note that, in the su(2) case, no such symmetric tensor appears,
and the spin-$\frac12$ Heisenberg model with
alternating exchange coupling is reduced to the spin-1 Heisenberg model
in the limit of $J'\to-\infty$.
To cancel the additional contributions from $d_{abc}$ in the su(3) case,
we need to modify $\mathcal{H}_3$ in
Eq.\ (\ref{eq: S=1 Hamiltonian with staggered biquadratic terms})
to the following form:
\begin{align}
\widetilde{\mathcal{H}}_3
=& -J' \sum_{i} (\Vec{S}_{i,1} \cdot \Vec{S}_{i,2})^2 \n
&+\frac{J}{2} \sum_{i,a}
\left[\hat\lambda_{a(i,1)} + \hat{\bar{\lambda}}_{a(i,2)}\right]
\left[\hat\lambda_{a(i+1,1)} + \hat{\bar{\lambda}}_{a(i+1,2)}\right],\n
=& -J' \sum_{i} (\Vec{S}_{i,1} \cdot \Vec{S}_{i,2})^2 \n
&+J \sum_i\left[-(\Vec{S}_{i,1} \cdot \Vec{S}_{i+1,2})^2
-(\Vec{S}_{i,2} \cdot \Vec{S}_{i+1,1})^2 \right.\n
&\qquad\qquad
+(\Vec{S}_{i,1} \cdot \Vec{S}_{i+1,1})^2 +\Vec{S}_{i,1} \cdot \Vec{S}_{i+1,1}\n
&\left.\qquad\qquad
+(\Vec{S}_{i,2} \cdot \Vec{S}_{i+1,2})^2 +\Vec{S}_{i,2} \cdot \Vec{S}_{i+1,2}
\right],
\end{align}
where $\hat{\lambda}_{a(i,n)}$ and $\hat{\bar{\lambda}}_{a(i,n)}$ are
the $\hat{\lambda}_a$ and $\hat{\bar{\lambda}}_a$ operators on
the site $(i,n)$.
In the limit $J'/J \to -\infinity$,
this Hamiltonian reduces to the SU(3) Hamiltonian $H_\theta$ at $\theta=0$
given in Eq.~(\ref{eq: SU(3) Hamiltonian}).
As shown in Sec.\ \ref{sec: DMRG}, the ground state of $H_\theta$ at $\theta=0$
is in the same $\mathbb{Z}_3$ SPT phase as the ground state of
the SU(3) AKLT model.
We note that the Hamiltonian $\mathcal{H}_3$ 
is obtained by dropping several terms proportional to $J$
in $\widetilde{\mathcal{H}}_3$
and has a simpler form in $S=1$ spin operators.
The iDMRG result for $\mathcal{H}_3$ shows that 
the nature of the $\mathbb{Z}_3$ SPT phase 
is not destroyed
even with this simplification of the Hamiltonian.

\subsubsection{$Z_3 \times Z_3$ symmetry}

As we discussed in Sec.\ \ref{sec: SU(3) AKLT model},
the $\mathbb{Z}_3$ SPT phase protected by $Z_3 \times Z_3$ symmetry
is characterized by projective representations of the symmetry.
For the  the $S=1$ spin chains we have introduced above,
the symmetry operations (in the linear representation)
are defined for two spins in the unit cell,
\begin{align}
\begin{aligned}
x&=\exp\left\{ \frac{2\pi i}{3}\!
\left[\frac{(\hat \lambda_2+\hat{\bar{\lambda}}_2)
-(\hat \lambda_5+\hat{\bar{\lambda}}_5)
+(\hat \lambda_7+\hat{\bar{\lambda}}_7)}{\sqrt{3}} \right]\right\},
\\
y&=\exp\left\{ \frac{2\pi i}{3}\!
\left[\frac{(\hat \lambda_3+\hat{\bar{\lambda}}_3)
     -\sqrt{3} (\hat \lambda_8+\hat{\bar{\lambda}}_8)}{2} \right]
\right\}.
\end{aligned}
\end{align}
Since both models, $\mathcal{H}_3$ and $\widetilde{\mathcal{H}}_3$,
have $S=1$ spins in $\bm 3$ and $\bm\bar{\bm 3}$ representations
in the unit cell,
the projective representation of the symmetry group is readily seen
as symmetry operations for individual $S=1$ spins.
Namely, the operation of $U_x$ interchanges three states
$\ket{x},\ket{y},\ket{z}$,
while the operation of $U_y$ gives different U(1) phase factors to
$\ket{x},\ket{y},\ket{z}$.

\subsection{SU(4) AKLT model}

We are going to argue that a variant of the SU(4) symmetric
Kugel-Khomskii model\cite{kugel-khomskii82} has a dimerized ground state
which is adiabatically connected to the AKLT state of
a $\mathbb{Z}_4$ SPT phase.
In a similar way to the case of the SU(3) AKLT model,
we can obtain the SU(4) AKLT model in the strong-coupling limit.

We begin with a review of
the symmetric Kugel-Khomskii model\cite{kugel-khomskii82,lundgren12}
which has two spin and two orbital degrees of freedom at each site
represented by two sets of Pauli matrices
$s^\alpha$ and $\tau^\beta$:
\begin{align}
\mathcal{H}_\mathrm{KK}= J \sum_i
\left( \bm{s}_i\cdot\bm{s}_{i+1}+ 1 \right)\!
\left( \bm{\tau}_i\cdot\bm{\tau}_{i+1}+ 1 \right).
\label{eq: H Kugel-Khomskii}
\end{align}
It apparently has SU(2)$\times$SU(2) symmetry in spin and orbital
spaces and is invariant under exchanging $\bm{s}$ and $\bm{\tau}$.
It is well known that the model has actually larger SU(4)
symmetry,\cite{Li-prl98,yamashita98}
as we briefly review below.
On each site we have four states $\ket{s^z,\tau^z}$,
which we label as
\begin{align}
\label{4 states}
\begin{aligned}
\ket{1}&=\ket{\mbox{$+1,+1$}}, & \ket{2}&=\ket{\mbox{$-1,+1$}}, \\
\ket{3}&=\ket{\mbox{$+1,-1$}}, & \ket{4}&=\ket{\mbox{$-1,-1$}}.
\end{aligned}
\end{align}
They form basis states for the fundamental representation
$\bm{4}$ of su(4),
in which 15 generators of su(4) $A^a$ ($a=1,\ldots,15$)
are given by
\begin{align}
s^\alpha, \tau^\alpha, s^\alpha \tau^\beta, \quad (\alpha,\beta=x,y,z).
\end{align}
We note that we have adopted the normalization of the su(4) generators
as
$\t{tr}(A^a A^b)=4\delta_{ab}$,
which differs from the normalization of the su(3) generators
in the previous sections.
Since the Casimir operator for the $\bm 4 \tensor \bm 4$
representation formed by the states at the $i$th and $j$th sites is given by
\begin{equation}
\sum^{15}_{a=1} A^a_i A^a_j
=\left(\bm{s}_i\cdot\bm{s}_j+ 1 \right)\!
\left(\bm{\tau}_i\cdot\bm{\tau}_j+ 1 \right)-1,
\end{equation}
the Hamiltonian in Eq.~(\ref{eq: H Kugel-Khomskii})
can be written as a sum of Casimir operators, and
therefore it has global SU(4) symmetry.

Following the discussion for the SU(3) case
in Sec.\ \ref{sec: spin quadrupolar},
let us consider a 1D lattice with the unit cell containing two sites,
one with the fundamental representation $\bm 4$ and
the other with its conjugate representation $\bm\bar{\bm 4}$.
Our idea is to make use of the quadratic Casimir operator
for $\bm 4 \tensor \bm\bar{\bm 4}$ representation from neighboring sites
to design a ground-state wave function which resembles
an MPS with $\mathbb{Z}_4$ SPT order.
The su(4) generators in the $\bm\bar{\bm 4}$ representation $\tilde A^a$
are given by
\begin{align}
\tilde A^a=-(A^a)^*.
\end{align}
Thus the Casimir operator for $\bm 4 \tensor \bm\bar{\bm 4}$ representations
is
\begin{align}
\sum^{15}_{a=1} A^a_i \tilde A^a_j
=
-&\left(s_i^x s_j^x -s_i^y s_j^y+s_i^z s_j^z+ 1 \right)
 \n
\times&
\left(\tau_i^x \tau_j^x -\tau_i^y \tau_j^y
         +\tau_i^z \tau_j^z + 1 \right) + 1,
\end{align}
which, unfortunately, is less symmetric and
conserves neither $s^z$ nor $\tau^z$.
However, we can perform a unitary transformation in
the $\bm\bar{\bm 4}$ representation,
\begin{align}
\tilde A^a \to (s^y \tau^y)\tilde A^a (s^y \tau^y),
\end{align}
to transform the Casimir operator back to the form
\begin{equation}
\sum^{15}_{a=1} A^a_i \tilde A^a_j
=-\left(\bm{s}_i\cdot\bm{s}_j - 1 \right)\!
\left( \bm{\tau}_i\cdot\bm{\tau}_j - 1 \right)
+1,
\label{Casimir SU(4)}
\end{equation}
which manifestly recovers the SU(2)$\times$SU(2) symmetry.

The product of the $\bm 4$ and $\bm\bar{\bm 4}$ representations
from neighboring sites is decomposed as
\begin{align}
\bm 4 \tensor \bm\bar{\bm 4} = \bm{15} \oplus \bm 1.
\end{align}
The eigenvalue of the Casimir operator in Eq.\ (\ref{Casimir SU(4)})
is $C(\bm{15})=4$ and $C(\bm{1})=0$.
As we have discussed for the SU(3) case in Sec.\ \ref{sec: SU(3) AKLT model B},
the MPS wave function of AKLT type which has $\mathbb{Z}_4$ SPT order
is obtained by projecting the $\bm 4 \tensor \bm\bar{\bm 4}$ states
from two neighboring sites onto $\bm{15}$ and $\bm 1$ in alternating
order along the 1D lattice.
This motivates us to consider the spin-orbital model with alternating
sign of coupling,
\begin{align}
\mathcal{H}_4=&
-J'\sum_i
\left( \bm{s}_{i,1}^{}\cdot\bm{s}_{i,2}^{} - 1 \right)\!
\left( \bm{\tau}_{i,1}^{}\cdot\bm{\tau}_{i,2}^{} - 1 \right) \n
&
-J\sum_i
\left( \bm{s}_{i,2}^{}\cdot\bm{s}_{i+1,1}^{} - 1 \right)\!
\left( \bm{\tau}_{i,2}^{}\cdot\bm{\tau}_{i+1,1}^{} - 1 \right),
\end{align}
where the spin-orbital exchange on each bond favors either
$\bm{15}$ or $\bm 1$ state depending on the sign of 
the coupling $J'$ or $J$.
In view of the numerical results for the similar model for the
SU(3) case in Eq.\ (\ref{eq: S=1 Hamiltonian with staggered biquadratic terms}),
we expect that the ground state of this Hamiltonian for $J'<0$
and $J>0$ should be adiabatically connected to the dimerized state
where a singlet is formed on every bond connecting neighboring unit cells
and four states in Eq.\ (\ref{4 states}) are left as zero-energy
end states when the 1D lattice is cut between two unit cells.

Finally, we propose a model which reduces to the SU(4) bilinear
exchange Hamiltonian in the limit of strong coupling.
The Hamiltonian is given by
\begin{align}
\widetilde{\mathcal{H}}_4=&
-J'\sum_i
\left( \bm{s}_{i,1}\cdot\bm{s}_{i,2} - 1 \right)\!
\left( \bm{\tau}_{i,1}\cdot\bm{\tau}_{i,2} - 1 \right) \n
& +J\sum_{i,a}
(A^a_{i,1} + \tilde A^a_{i,2} )
(A^a_{i+1,1} + \tilde A^a_{i+1,2} ),
\end{align}
with $J>0$ and $J'<0$.
In the limit $J'\to -\infinity$, we have only the $\bm{15}$ representation
in each unit cell.
The effective Hamiltonian for the interaction between neighboring
$\bm{15}$ representations, which can be obtained in the same way as
in Sec.\ \ref{sec: spin quadrupolar}, has the form
\begin{equation}
\mathcal{H}_{\mathrm{eff}}=J\sum_i\sum_{a=1}^{15}A^a_i A^a_{i+1},
\end{equation}
where $A^a_i$ ($a=1,\ldots,15$) are the generators 
of su(4) in the $\bm{15}$ (adjoint) representation.
In view of the fact that the SU(2) and SU(3) versions of the Hamiltonian
have the ground state with SPT order [$S=1$ Haldane phase for SU(2)
and the numerical result in Sec.\ \ref{sec: DMRG} for SU(3)],
we expect that the ground state of this Hamiltonian should be in
the $\mathbb{Z}_4$ SPT phase protected by $Z_4\times Z_4$ symmetry,
as in the SU(4) AKLT model.
We note that the above Hamiltonian naturally has the $Z_4 \times Z_4$ symmetry
which is a subgroup of the SU(4) symmetry.

\section{Summary \label{sec: summary}}
We have studied $\mathbb{Z}_3$ SPT phases protected by
global $Z_3\times Z_3$ symmetry.
By applying the group cohomology classification of 1D SPT phases
and using nontrivial cocycles of $H^2(Z_3\times Z_3)$,
we have constructed MPS wave functions of $\mathbb{Z}_3$ SPT phases,
which are SU(3) extensions of the AKLT wave function.
The MPS wave functions are ground states of 
the SU(3) bilinear-biquadratic Hamiltonian at $\theta=\arctan(2/9)$
[the SU(3) AKLT model].

Using the iDMRG method,
we have determined the phase diagram of
the SU(3) bilinear-biquadratic Hamiltonian,
which has the $\mathbb{Z}_3$ SPT phase and the dimer phase.
These phases are characterized by
an SU(3) version of the string order parameters
and dimerization, respectively.
The critical point separating the two phases is
located at $\theta_c \approx -0.027\pi$.
From the scaling of the entanglement entropy against the correlation length
we have obtained a central charge $c=16/5$ at the critical point,
suggesting that the criticality is described by the SU(3)$_2$ WZW model.

We have pointed out that the SU(3) bilinear Hamiltonian $H_{\theta=0}$
might be realized in the SU(3) Hubbard model
of two orbitals of fermions in the strong $U$ limit.
When one orbital is tuned to be $1/3$ filled and 
the other to be $2/3$ filled,
the charge and orbital sectors will be gapped, and
the low-energy effective model of the two-orbital SU(3) Hubbard model
will be an SU(3) spin chain in the adjoint representation.
We speculate that such a system might be realized with cold atoms.

We have proposed $S=1$ spin chains with staggered biquadratic couplings 
that are adiabatically connected to the SU(3) AKLT model.
In view of a proposal of realizing the $S=1$ bilinear-biquadratic model
using cold atoms,\cite{garcia-ripoll04}
we consider that our $S=1$ spin model might also be realized in cold atoms
by properly engineering staggered biquadratic couplings.

We have also proposed a variant of Kugel-Khomskii model
with spin-1/2 and two orbital degrees of freedom
which is connected to the SU(4) AKLT model.

\begin{acknowledgments}
It is our pleasure to acknowledge stimulating discussions with Masaki Oshikawa.
This work was supported by Grants-in-Aid from the Japan Society for
Promotion of Science (Grants No.~24840047, No.~25800221, No.~23540397 and No.~24540338)
and by the RIKEN iTHES Project.
\end{acknowledgments}

\appendix
\section{Group cohomology \label{app: group cohomology}}
Here we briefly review the group cohomology
of a group $G$ over $U(1)$.
We first define $n$-cochains $\phi^n \in C_n$ which are
functions from $G^n$ to $U(1)$,
\begin{equation}
\phi^n: G^n \to U(1).
\end{equation}
The set of $n$-cochains is denoted by $C_n$.
Then we define coboundary operators $\delta^n$
which transform $n$-cochains to $(n+1)$-cochains,
\begin{equation}
\delta^n: C_n \to C_{n+1},
\end{equation}
through the formula
\begin{align}
&(\delta^n \phi^n)(g_1,\dots,g_{n+1}) \n
&:=
\phi^n(g_2,\ldots,g_{n+1}) \n
&\hphantom{:=}
+\sum_{i=1}^n (-1)^i
 \phi^n(g_1,\ldots,g_{i-1},g_i g_{i+1},g_{i+2},\ldots,g_{n+1}) \n
&\hphantom{:=} +(-1)^{n+1}\phi^n(g_1,\ldots,g_n),
\label{eq: coboundary op}
\end{align}
where $g_1,\ldots,g_n,g_{n+1} \in G$.
Here we have assumed that actions of elements of $G$ on $U(1)$
are trivial,
i.e., $g_i$'s are unitary operators.
\footnote{Equation (\ref{eq: coboundary op}) needs to be
amended when antiunitary operators
such as time-reversal transformation $T$ are involved,
because an operation of $T$ changes the sign of $U(1)$ variable.
In that case
the first term on the right hand side of Eq.~(\ref{eq: coboundary op})
should be acted by $g_1$.
However, we consider only unitary symmetries throughout this paper.}
The identity
\begin{equation}
\delta^{n+1} \circ \delta^n=0
\end{equation}
holds.
We have a sequence of homomorphisms (cochain complex)
from coboundary operators,
\begin{align}
0 \to C_0 \to C_1 \to \cdots \to C_{n-1} \to C_n \to C_{n+1} \to \cdots,
\end{align}
and we define cohomology groups for the above cochain complex as
\begin{align}
H^n(G,U(1))=\t{Ker}\, \delta^n/\t{Im}\, \delta^{n-1}.
\label{eq: group cohomology}
\end{align}
Here $Z_n=\t{Ker}\, \delta^n$ is called $n$-cocycles,
and $B_n=\t{Im}\, \delta^{n-1}$ is $n$-coboundaries.
Let us write down conditions for $Z_2$ and $B_2$ explicitly.
A 2-cocycle $\phi^2 \in Z_2$ satisfies
\begin{align}
\phi^2(g_2, g_3)-\phi^2(g_1 g_2, g_3)+\phi^2(g_1,g_2 g_3)-\phi^2(g_1, g_2)=0.
\label{eq: 2-cocycle app}
\end{align}
A 2-coboundary $\phi^2 \in B_2$ is obtained from
a 1-cochain $\phi^1\in C_1$ as
\begin{align}
\phi^2(g_1,g_2)=\phi^1(g_2) -\phi^1(g_1 g_2)+\phi^1(g_1).
\label{eq: 2-coboundary app}
\end{align}
The phase functions $\phi(g_1,g_2)$ that appeared in symmetry
transformations of MPSs in Sec.~\ref{sec:MPS and group cohomology}
satisfy the consistency condition [Eq.~(\ref{eq: 2 cocycle condition})]
and the equivalence relation [Eq.~(\ref{eq: coboundary equivalence})].
The former coincides with the 2-cocycle condition
of Eq.\ (\ref{eq: 2-cocycle app}),
while the latter means the equivalence up to 2-coboundaries.
Thus the phase functions $\phi(g_1,g_2)$ in the symmetry
transformations of MPSs are elements of the
second cohomology group $H^2(G,U(1))$.

\section{Group cohomology for $G=Z_N\times Z_N$ \label{app: Z_N * Z_N}}
We summarize results for the second cohomology group of the group
$G=Z_N\times Z_N$ over U(1), which is obtained
by applying the K\"{u}nneth formula and the universal coefficient
theorem.\cite{hatcher-AT,chen-gu-liu-wen13}
We write down a non-trivial cocycle for $H^2(Z_N\times Z_N,U(1))$.

\subsection{K\"{u}nneth formula}
The universal coefficient theorem indicates an isomorphism
\begin{align}
H^n(G,M)=\t{Hom}[H_n(G,\mathbb{Z}),M]\oplus \t{Ext}[H_{n-1}(G,\mathbb{Z}),M],
\end{align}
for the cohomology over an Abelian group $M$ and
the homology over $\mathbb{Z}$.
Since the
$\t{Ext}$ functor vanishes
($\t{Ext}[Z,U(1)]=0$ and $\t{Ext}[Z_m,U(1)]=0$)
for $M=U(1)$,
we have
\begin{align}
H^n(G,U(1))=\t{Hom}[H_n(G,\mathbb{Z}),U(1)].
\end{align}
For discrete group $G$,
$H^n(G,U(1))$ for $n\ge 1$ has a torsion part only
and is given by
\begin{align}
H^n(G,U(1))=H_n(G,\mathbb{Z})
\label{eq: H^n over U(1) and H_n over Z}
\end{align}
for $n\ge 1$.
From Eq.~(\ref{eq: H^n over U(1) and H_n over Z}), we have
\begin{align}
\begin{aligned}
H_1(Z_N,\mathbb{Z})&=H^1(Z_N,U(1))=\mathbb{Z}_N, \\
H_2(Z_N,\mathbb{Z})&=H^2(Z_N,U(1))=0.
\end{aligned}
\end{align}
The first line comes from the formula $H^1(G,U(1))=G$
for Abelian group $G$,
while the second line is obtained from an explicit calculation
of $H^2(Z_N,U(1))$ from Eq.~(\ref{eq: group cohomology}).
The zeroth homology group is known to be given by
\begin{align}
H_0(Z_N,\mathbb{Z})=\mathbb{Z}.
\end{align}

Homology groups of the direct product $G=G_1 \times G_2$
of groups $G_1$ and $G_2$
can be computed with the use of the K\"{u}nneth formula\cite{hatcher-AT}
that gives the isomorphism
\begin{align}
H_n(G_1 \times G_2, \mathbb{Z})=&
\bigoplus_i H_i(G_1,\mathbb{Z}) \otimes H_{n-i}(G_2,\mathbb{Z}) \n
&\oplus
\bigoplus_i \t{Tor}[H_i(G_1,\mathbb{Z}),H_{n-i-1}(G_2,\mathbb{Z})].
\label{kunneth formula}
\end{align}
Therefore, we can obtain the second homology group
$H_2(Z_N \times Z_N, U(1))$ from the homology groups of $Z_N$.
Since the torsion functor vanishes
($\t{Tor}[\mathbb{Z},\mathbb{Z}_N]=\t{Tor}[\mathbb{Z}_N,\mathbb{Z}]=0$)
in Eq.\ (\ref{kunneth formula}),
we have
\begin{align}
H_2(Z_N \times Z_N, \mathbb{Z})
=H_1(Z_N , \mathbb{Z}) \tensor H_1(Z_N , \mathbb{Z}),
\end{align}
and finally we obtain the second cohomology group of $G=Z_N\times Z_N$,
\begin{align}
H^2(Z_N \times Z_N, U(1))
&=H_1(Z_N , U(1))\tensor H_1(Z_N , U(1))  \n
&= \mathbb{Z}_N \tensor \mathbb{Z}_N \n
&= \mathbb{Z}_N.
\label{eq: H^2 for G=Z_N*Z_N}
\end{align}

\subsection{Non-trivial 2-cocycles of $H^2(Z_N \times Z_N, U(1))$}
The nontrivial cocycle of
$H^2(Z_N \times Z_N, U(1))=\mathbb{Z}_N$ is found from the isomorphism
\begin{align}
\t{Hom}[H_1(Z_N,\mathbb{Z}) &\otimes H_1(Z_N,\mathbb{Z}),U(1)] \n
&\to H^2(Z_N \times Z_N, U(1)).
\label{eq: H^2 isomorphism}
\end{align}
Let us denote the group elements of $Z_N \times Z_N$ by
\begin{subequations}
\label{eq: group element of Z_N * Z_N}
\begin{align}
x^{n_1}y^{n_2}
\qquad
(n_1,n_2=0,\ldots,N-1),
\end{align}
where $x$ and $y$ are generators of the first and the second $\mathbb{Z}_N$,
respectively, satisfying
\begin{equation}
x^N=1, \qquad y^N=1.
\end{equation}
\end{subequations}
We can define a set of 2-cocycles $\phi^2=m \varphi$,
\begin{subequations}
\label{eq: 2 cocycle ZN times ZN}
\begin{align}
\phi^2=m \varphi : (Z_N \times Z_N)^2 \to U(1)
\end{align}
for $m=0,\ldots,N-1$, where
\begin{align}
m \varphi(x^{n_1}y^{n_2},x^{n'_1}y^{n'_2})= n_1 n'_2 m\frac{2\pi}{N}
\quad\mathrm{mod}\;2\pi.
\end{align}
\end{subequations}
We have constructed this 2-cocycle from the isomorphism
in Eq.~(\ref{eq: H^2 isomorphism}) using the following sequence of
mappings:
\begin{equation}
\begin{matrix}
(Z_N \times Z_N)^2 &\to& Z_N \times Z_N &\to& Z_N &\to& U(1), \\
(x^{n_1}y^{n_2},x^{n'_1}y^{n'_2}) &\to& (x^{n_1},y^{n'_2}) &\to& n_1 n'_2 &\to&
\displaystyle n_1 n'_2 m\frac{2\pi}{N}.
\end{matrix}
\end{equation}
The functions $m\varphi$ clearly satisfy the 2-cocycle condition
[Eq.~(\ref{eq: 2-cocycle app})] as
\begin{align}
&\delta(m\varphi)(x^{n_1}y^{n_2},x^{n'_1}y^{n'_2},x^{n''_1}y^{n''_2}) \n
&=m \frac{2\pi}{N}[n'_1 n''_2 - (n_1+n'_1)n''_2 + n_1(n'_2+n''_2)-n_1 n'_2] \n
&=0.
\end{align}
They are not 2-coboundaries [Eq.~(\ref{eq: 2-coboundary app})] except
the one with $m=0$,
because they have the property
$m\varphi(x,y) \neq m\varphi(y,x)$ for $m=1,\ldots,N-1$,
whereas 2-coboundaries must satisfy the relation
\begin{align}
\delta \phi^1(x,y)=\delta \phi^1(y,x).
\end{align}
Thus $\phi^2=m\varphi$ ($m=0,\ldots,N-1$) form an Abelian group $\mathbb{Z}_N$
with addition of functions and give nontrivial 2-cocycles for $m=1,\ldots,N-1$.
Clearly, $\varphi$ is a generator of the cohomology group 
$H^2(Z_N \times Z_N, U(1))=\mathbb{Z}_N$.
In the case of $N=3$, we have
\begin{align}
\exp[i\varphi (x^{n_1} y^{n_2}, x^{n'_1} y^{n'_2})]
= \omega^{n_1 n'_2}
\end{align}
with $\omega=\exp(2\pi i/3)$.
In particular,
\begin{align}
\exp[i\varphi (x,y)]&=\omega, &
\exp[i\varphi (y,x)]&=1.
\end{align}
We can use elements of the second cohomology group $H^2(Z_3\times Z_3,U(1))$
to construct projective representations of $G=Z_3\times Z_3$.
With the 2-cocycle $\phi^2=\varphi$ we find from
Eq.\ (\ref{projective representation}) that generators $U_x$ and $U_y$
of a projective representation obey
\begin{align}
U_x U_y= \omega U_y U_x
\end{align}
and do not commute, as opposed to the elements
$x$ and $y$ of the group $Z_3\times Z_3$.

\section{SU($N$) AKLT states and their transfer matrices \label{app: SU(N) AKLT}}

We show several properties of the transfer matrix $\mathcal{M}$
that are used in calculating correlation functions
in Sec.~\ref{subsec: SU(N) AKLT}.
We consider the SU($N$) AKLT wave function of the MPS form
\begin{align}
\ket{\Psi}&=C_N^{-1}\sum_{ \{\sigma_i \} } 
\t{tr}[A^{\sigma_1}A^{\sigma_2} \ldots A^{\sigma_L}]
\ket{\sigma_1 \sigma_2 \ldots \sigma_L},
\end{align}
with 
\begin{align}
A^\sigma&=\sqrt{\frac{2}{N}} t^\sigma,
\end{align}
where $C_N$ is a normalization constant,
$\sigma_i$ labels states on each site in the adjoint ($\bm{N^2-1}$)
representation,
and $t^a$ ($a=1,\ldots,N^2-1$) are the su($N$) generators
in the fundamental representation.
The su($N$) generators $t^a$ are traceless and hermitian $N\times N$ matrices
that are normalized as
\begin{align}
\t{tr}(t^a t^b)=\frac{1}{2}\delta_{ab}
\end{align}
and obey the commutation relations 
\begin{align}
[t^a,t^b]=if_{abc}t^c.
\end{align}
Here the structure constants $f_{abc}$ are totally antisymmetric,
and summation over the repeated index $c$ is assumed.
The quadratic Casimir operator of su($N$) operators $T^a$ in 
the $\bm{d}$-dimensional representation is written as
\begin{align}
\sum_{a=1}^{N^2-1}T^a T^a= C(\bm{d}) 1_d.
\end{align}
For the fundamental representation ($T^a_{jk}=t^a_{jk}$)
and the adjoint representation ($T^a_{jk}=-if_{ajk}$),
the eigenvalue of the quadratic Casimir operator reads
\begin{align}
C(\bm{N})&=\frac{N^2-1}{2N}, & C(\bm{N^2-1})&=N.
\end{align}
In addition, we have a formula \cite{DiFrancesco}
\begin{align}
\t{tr}(T^a T^b)=\frac{d C(\bm{d})}{N^2-1} \delta_{ab}.
\label{eq: tr ta tb}
\end{align}

Next we define a transfer matrix 
for the SU($N$) AKLT state as
\begin{align}
\mathcal{M}=\sum_{m=1}^{N^2-1} A^m \tensor (A^m)^*,
\end{align}
which is an $N^2\times N^2$ matrix.
The $N^2$-dimensional vector space is spanned by 
a basis $\{ \ket{i}\tensor \ket{j} \}$ with $i,j=1,\ldots,N$,
where $\{ \ket{i} \}$ is an orthonormal basis of
the $N$-dimensional vector space.
For two $N^2$-dimensional vectors
\begin{align}
\ket{u}&=\sum_{ij} u_{ij}\ket{i}\tensor \ket{j}, &
\ket{v}&=\sum_{ij} v_{ij}\ket{i}\tensor \ket{j},
\end{align}
the inner product is written as 
\begin{align}
\langle u \ket{v}=\t{tr}(u^\dagger v).
\end{align}
With this basis, the action of the transfer matrix reads
\begin{align}
\mathcal{M}\ket{u}&=\sum_{m,i,j}
[A^m u (A^m)^\dagger]_{ij} \ket{i}\tensor \ket{j} .
\end{align}

Now we show Eq.~(\ref{eq: eigs of M}). 
This follows from the fact that an orthonormal basis
of the $N^2$-dimensional vector space
\begin{align}
\ket{v_0}&= \frac{1}{\sqrt{N}} \sum_i \ket{i} \tensor \ket{i}, \\
\ket{v_a}&= \sqrt{2} \sum_i t^a_{ij} \ket{i} \tensor \ket{j}, 
\end{align}
is a set of all eigenvectors of the transfer matrix $\mathcal{M}$ satisfying
\begin{align}
\mathcal{M} \ket{v_0}&=  \frac{N^2-1}{N^2} \ket{v_0}, &
\mathcal{M} \ket{v_a}&= -\frac{1}{N^2} \ket{v_a}.
\end{align}
This can be seen by using the following three equations:
\begin{align}
\bra{v_0}\mathcal{M}\ket{v_0}
&= \frac{2}{N^2} \sum_a \t{tr}(t^a t^a) 
=\frac{N^2-1}{N^2}, \\
\bra{v_a}\mathcal{M}\ket{v_0}
&= \left(\frac{2}{N} \right)^\frac{3}{2} \sum_{a'} \t{tr}(t^a t^{a'} t^{a'})
=0, 
\end{align}
and 
\begin{align}
\bra{v_a}\mathcal{M}\ket{v_b}
&=
\frac{4}{N} \sum_{c} \t{tr}(t^a t^c t^b t^c) \n
&=\frac{4}{N} \sum_{c} 
\left[\t{tr}(t^a t^c t^c t^b) + \sum_d if_{bcd}\t{tr}(t^a t^c t^d) \right] \n
&=\frac{4}{N} 
\left[\frac{C(\bm{N})}{2}\delta_{ab}
-\frac{1}{4}\sum_{c,d}f_{bcd}f_{cda} \right]
\n
&=-\frac{1}{N^2}\delta_{ab}.
\end{align}
In the last equation,
we have used Eq.~(\ref{eq: tr ta tb}) for the adjoint representation 
in which $T^a_{bc}=-if_{abc}$.

Finally, we show Eq.~(\ref{eq: orthogonality of v0 and Ma v0}).
In our basis, the action of $\widetilde{\mathcal{M}}^a$ reads
\begin{align}
\widetilde{\mathcal{M}}^a \ket{u}
&=\sum_{m,n,i,j}
-if_{amn} [A^n u (A^m)^\dagger]_{ij} \ket{i}\tensor \ket{j} .
\end{align}
Then Eq.~(\ref{eq: orthogonality of v0 and Ma v0}) follows as
\begin{align}
\bra{v_0} \widetilde{\mathcal{M}}^a \ket{v_0}
&=-\frac{2i}{N^2} \sum_{m,n} f_{amn} \t{tr}(t^n t^m)=0,
\end{align}
because $f_{abc}$ is a totally antisymmetric tensor.
In view of Eq.~(\ref{eq: eigs of M}),
this implies that $\widetilde{\mathcal{M}}^a \ket{v_0}$
is an eigenvector of $\mathcal{M}$ with the eigenvalue $-1/N^2$.

%

\end{document}